%% file: main.tex
\documentclass[letterpaper,journal]{IEEEtran}

\usepackage{cite}
\usepackage{textcomp}
\usepackage{stfloats}
\usepackage{url}
\usepackage{balance}

\input{preamble}

\def\BibTeX{{\rm B\kern-.05em{\sc i\kern-.025em b}\kern-.08em
    T\kern-.1667em\lower.7ex\hbox{E}\kern-.125emX}}

\begin{document}

\title{Fingerprinting Text-to-Image Diffusion Models via Collapsed Generation}

\author{Yuanmin Huang, Chen Chen, Geng Hong, Xiaoyu You, Hui Xue, Zhenxing Qian, Mi Zhang, and Min Yang
\thanks{Yuanmin Huang, Chen Chen, Geng Hong, Zhenxing Qian, Mi Zhang, and Min Yang are with Fudan University, Shanghai, China (e-mail: \url{yuanminhuang23@m.fudan.edu.cn}; \url{chenc24@m.fudan.edu.cn}; \url{ghong@fudan.edu.cn}; \url{zxqian@fudan.edu.cn}; \url{mi_zhang@fudan.edu.cn}; \url{m_yang@fudan.edu.cn}).}
\thanks{Xiaoyu You is with East China University of Science and Technology, Shanghai, China (e-mail: \url{xiaoyuyou@ecust.edu.cn}).}
\thanks{Hui Xue is with Alibaba Group, Hangzhou, China (e-mail: \url{hui.xueh@alibaba-inc.com}).}
\thanks{Min Yang is also with Shanghai Pudong Research Institute of Cryptology, and Engineering Research Center of Cyber Security Auditing and Monitoring, Ministry of Education, China.}}

\maketitle

\input{tex/abstract}

\begin{IEEEkeywords}
Text-to-Image Diffusion Models, Intellectual Property Protection, Model Fingerprinting, Ownership Verification.
\end{IEEEkeywords}

\input{figs/intro_fig}

\input{tex/intro}
\input{tex/related_work}

\input{tex/prelim}

\input{tex/method}

\input{tex/experiment}

\input{tex/conclusion}

\bibliographystyle{IEEEtran}
\bibliography{main}

\clearpage
\appendices

\input{tex/supplements}

\end{document}

%% file: preamble.tex
\usepackage{tikz}
\usepackage{amsfonts}
\usepackage{amsmath}
\usepackage{amssymb}
\usepackage{amsthm}
\usepackage{bm}
\usepackage{booktabs}
\usepackage{graphicx}
\usepackage{siunitx}
\usepackage{comment}
\usepackage{hyperref}

\usepackage{multirow} \usepackage{multicol}

\usepackage{tabularx} \usepackage{array, colortbl} 
\usepackage{caption}  \usepackage{subcaption}

\newtheorem{definition}{Definition}

\PassOptionsToPackage{table,xcdraw}{xcolor}
\usepackage{xcolor}

\definecolor{TablePurple}{RGB}{151, 107, 185}
\definecolor{TableBlue1}{RGB}{78, 98, 171}   \definecolor{TableBlue2}{RGB}{70, 158, 180}  \definecolor{TableGreen}{RGB}{135, 207, 164} \definecolor{TableYellow}{RGB}{254, 232, 154}

\usepackage{algorithm}
\usepackage{algorithmicx}
\usepackage{algpseudocode}
\usepackage[switch]{lineno}

\usepackage{bbding}

\usepackage{enumitem}
\setlist[itemize]{leftmargin=*}

%% file: tex/abstract.tex
\begin{abstract}

Proprietary text-to-image diffusion models are increasingly distributed as hosted services and downloadable checkpoints, making their intellectual property (IP) protection an increasingly critical concern when model leakage, copying, or unauthorized fine-tuning is disputed.
In this work, we present a non-invasive model fingerprinting framework based on \emph{collapsed generation}, a phenomenon where certain input conditions produce highly consistent images across multiple stochastic seeds.
We show that collapsed generation is an intrinsic, model-dependent property of the learned generation process.
These collapse-prone conditions therefore expose model-specific behavioral signatures, enabling reliable ownership verification without embedding invasive watermarks.
After preparing conditions on the source model, the framework verifies a suspect model under two access settings: (1) white-box pipeline access, where optimized continuous embeddings can be injected into the generation process, and (2) black-box API-only access, where natural language prompts are queried through the service interface.
In both cases, ownership evidence is measured by whether the suspect model reproduces the source model's collapse behavior across stochastic samplings.
Extensive experiments across UNet- and transformer-based diffusion models show that collapsed generation fingerprints can distinguish different source models with low confusion.
These fingerprints remain verifiable in fine-tuned derivatives and under common and adaptive model- or query-level obfuscations, while requiring only a modest verification query budget.
Together, these results establish collapsed generation as a reliable intrinsic evidence source for non-invasive diffusion model ownership verification.

\end{abstract}

%% file: figs/intro_fig.tex
\begin{figure*}[!t]
    \centering
    \includegraphics[width=0.98\linewidth]{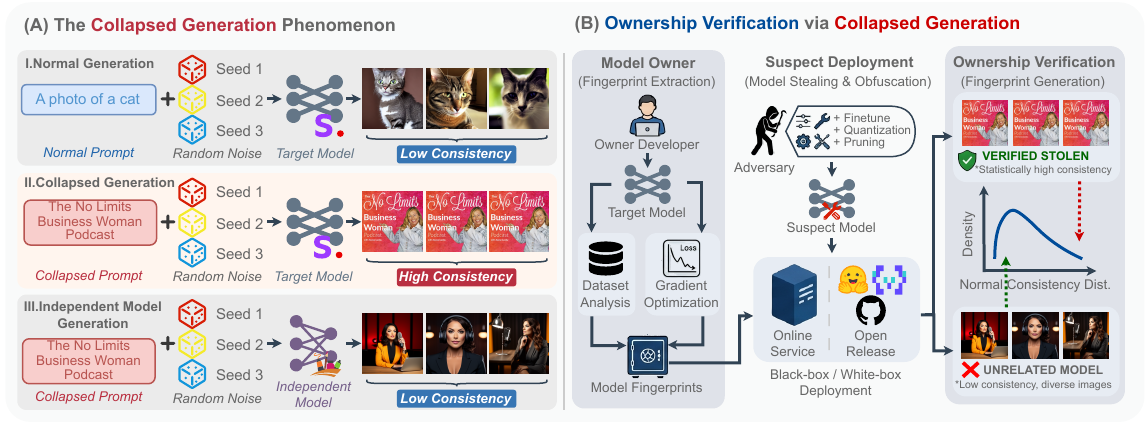}
    \caption{\textbf{Collapsed generation as a non-invasive ownership signal for diffusion models.}
    \textbf{(a) Collapsed generation.} Under ordinary input conditions, different stochastic seeds usually produce diverse images, whereas a collapse-prone condition yields highly consistent outputs on the source model. Such collapse behavior is model-dependent and therefore can serve as a behavioral fingerprint.
    \textbf{(b) Ownership verification.} The model owner prepares such conditions on the source model, and the verifier tests whether a suspect model reproduces the corresponding cross-seed consistency. The framework supports both pipeline-access and API-only verification, and remains effective under model obfuscations.
    }
    \label{fig:intro}
\end{figure*}

%% file: tex/intro.tex
\section{Introduction}
The rapid advancement of text-to-image (T2I) diffusion models has profoundly transformed generative AI, enabling high-fidelity image synthesis directly from natural language prompts~\cite{rombach2022high,stabilityai_sd21_2022}.
These models now serve as foundational components in creative industries and commercial pipelines~\cite{DeciFoundationModels,li2024hunyuanditpowerfulmultiresolutiondiffusion,chen2024pixartalpha}.
As proprietary checkpoints are increasingly released, shared, fine-tuned, or deployed behind commercial APIs, disputes over model leakage, unauthorized reuse, and derivative ownership have become a practical intellectual property (IP) concern~\cite{jiang2023comprehensive,miura2024megex}.
This growing tension underscores the need for dedicated IP protection mechanisms for diffusion models.

Existing IP protection methods for diffusion models generally fall into two categories: \textit{invasive model watermarking}~\cite{zhao2023recipe,fernandez2023stable,liu2023watermarking} and \textit{non-invasive model fingerprinting}~\cite{zeng2024huref,guo2025one,teng2025fingerprinting,li2025diffip}.
The former embeds ownership signals by altering model parameters, training objectives, or generation outputs, which often introduces additional computational overhead and may compromise generation quality.
In contrast, non-invasive fingerprinting methods attempt to identify models through their inherent behaviors under specific inputs, without modifying model parameters or retraining, thereby preserving model integrity and usability.

However, existing non-invasive fingerprints only partially address the verification needs arising from leaked checkpoints, unauthorized derivatives, and hosted T2I services.
Methods based on internal representations or denoising trajectories can be useful when the verifier obtains a checkpoint or a controllable generation pipeline, but their verification evidence is tied to model-internal readouts or specialized sampling procedures~\cite{li2025diffip,teng2025fingerprinting}.
Such assumptions do not hold when the suspect model is exposed only as a hosted T2I service that returns final images.
Conversely, query-based methods are closer to the API-only setting, but existing designs often rely on optimized prompt suffixes whose linguistic abnormality and induced semantic deviation make the verification queries easier to distinguish as non-ordinary service interactions~\cite{guo2025one}.
These limitations motivate a different type of fingerprint signal: one whose evidence is expressed in the model's generation behavior, can be actively elicited when the generation pipeline is controllable, and can still be observed through natural prompts when only API outputs are available.

In this work, we exploit an intrinsic behavior of diffusion models, termed \emph{collapsed generation}, as a non-invasive ownership signature.
Under ordinary prompts, independent stochastic samplings usually produce diverse images.
Under certain model-dependent input conditions, however, the sampling stochasticity is suppressed and different generation trajectories converge to visually similar outputs~\cite{carlini2023extracting,hintersdorf2024finding}, as illustrated in Fig.~\ref{fig:intro} (a).
Such collapse-prone conditions are shaped by the learned data distribution, optimization dynamics, and model architecture~\cite{somepalli2023understanding,wen2024detecting}, and therefore can expose behavioral signatures that are difficult for independently trained models to reproduce.

Based on this observation, we formulate collapse-prone conditions as model-intrinsic fingerprints for ownership verification.
A model owner first constructs a fingerprint set on the source model with the resulting conditions as reference fingerprints, as illustrated in Fig.~\ref{fig:intro} (b).
During verification, the verifier applies these conditions to a suspect model and measures whether its independent stochastic generations reproduce the characteristic cross-seed consistency of the source model.
We consider two verification interfaces for the suspect model.
Under white-box pipeline access, the verifier can control the generation process and inject continuous prompt embeddings, whereas under black-box API-only access, the verifier submits text prompts and observes the final generated images.

In ownership disputes involving leaked checkpoints or unauthorized derivative models (e.g., through LoRA fine-tuning), the suspect model may be available as a controllable generation pipeline.
Such pipeline access allows the verifier to actively elicit model-specific collapsed generation by searching the continuous text embedding space.
However, directly searching this space can be computationally demanding, as it requires repeated differentiation through the iterative generation process.
We find that the collapse signature becomes distinguishable during the early denoising stage, well before the final image is produced.
This observation motivates a truncated optimization strategy that efficiently synthesizes reproducible fingerprint embeddings for checkpoint-level ownership verification.

In other cases, an unauthorized model may be deployed behind a hosted T2I service and exposed only through an end-to-end API.
Verification must therefore rely on text prompts and the resulting generated images.
Consequently, the fingerprint conditions must be realized as text prompts rather than optimized continuous embeddings.
The challenge is to identify prompts that intrinsically trigger model-specific collapse without exhaustively screening a vast prompt space through repeated generation.
We find that naturally collapsed prompts are strongly concentrated among training samples with unusually low loss.
This relationship provides an efficient way to mine natural fingerprint prompts from training dynamics for service-level ownership verification.

Despite their different construction mechanisms, both realizations rely on the same verification principle.
A suspect model is associated with the source model when the prepared fingerprint conditions reproduce statistically abnormal consistency across independent generations.
This shared criterion provides a unified behavioral basis for ownership verification.

Our main contributions are summarized as follows:
\begin{itemize}
    \item We identify collapsed generation as a model-intrinsic behavioral signature and formulate its abnormal consistency across stochastic generations as a non-invasive fingerprint signal for T2I model ownership verification.
    \item We develop a unified framework that bridges white-box checkpoint-level and black-box service-level ownership verification through the same collapsed-generation signature.
    Building on insights from early denoising behavior and training dynamics, the framework constructs continuous embeddings for controllable pipelines and natural prompts for hosted APIs.
    \item We evaluate the framework across multiple T2I model families spanning UNet- and transformer-based architectures.
    The results demonstrate clear discrimination among independently trained models, while the fingerprints remain detectable after fine-tuning, pruning, quantization, and adaptive query-time interventions with a modest verification query budget.
\end{itemize}

%% file: tex/related_work.tex
\section{Related Work}

\input{tables/fingerprint_comparison}

\subsection{Model Ownership Verification}
Model ownership verification aims to determine whether a suspect model originates from, or is derived from, a source model claimed by an owner.
Existing approaches generally establish ownership through model watermarks or model fingerprints.
Model watermarking proactively embeds a secret ownership signal into the protected model through its parameters, training objective, or trigger-response behavior~\cite{Uchida_2017,adi2018turningweaknessstrengthwatermarking,rouhani2018deepsignsgenericwatermarkingframework}.
Model fingerprinting instead characterizes a trained model using distinctive decision boundaries, adversarial responses, or internal behaviors, without modifying the protected model~\cite{lukas2019deep,cao2021ipguard,zheng2022dnn}.
For reliable ownership verification, the verification signal should remain reproducible after model obfuscations, such as fine-tuning, pruning, and quantization, while being sufficiently specific to distinguish the claimed source from independently trained or unrelated models.
The realization of these requirements depends on the architecture and deployment interface of the protected model, motivating specialized IP protection methods for diffusion models.

\subsection{Diffusion Model IP Protection}

\noindent
\textbf{Invasive Model Watermarking.}
A prominent line of diffusion model IP protection relies on watermarking, which deliberately implants a predefined ownership signal before deployment and later verifies whether the signal can be recovered from the model's generated outputs.
By embedding an owner-controlled signal that can be recovered through prescribed queries or output analysis, these methods provide explicit evidence for subsequent ownership verification.
Zhao et al.~\cite{zhao2023recipe} introduce watermark-bearing examples during training so that the resulting model reproduces designated watermark patterns, while Liu et al.~\cite{liu2023watermarking} use secret trigger prompts to activate predefined watermark responses.
Stable Signature~\cite{fernandez2023stable} follows a different route by fine-tuning the latent decoder so that generated images carry an invisible signature without requiring a special trigger at inference time.
More recent methods incorporate watermark conditioning or specialized training objectives to improve the persistence of ownership signals after downstream adaptation such as fine-tuning~\cite{min2024watermarkconditioneddiffusionmodelip,wang2025sleepermarkrobustwatermarkfinetuning}.
Despite their different embedding mechanisms, these methods require an ownership signal to be deliberately introduced through the training data, optimization procedure, or model components before release.
Such intervention introduces additional training and deployment overhead and may perturb the model's original output distribution, potentially affecting its normal generation behavior.
It also limits retrospective protection, since an already trained and unmarked model must be fine-tuned or retrained before watermark-based verification can be applied.

\noindent
\textbf{Non-Invasive Model Fingerprinting.}
The limitations of model watermarking motivate non-invasive model fingerprinting, which derives ownership evidence from behaviors already present in a trained model without modifying its parameters or training objective to implant an additional signal.
Existing diffusion model fingerprints can be organized primarily according to the verifier's access to the suspect model during ownership verification.
This access constrains both the fingerprint conditions that can be applied to the suspect model and the evidence available for the final ownership decision.
Note that white-box and black-box here refer to verification-time access to the suspect model rather than the owner's access to the source model during fingerprint construction.
Table~\ref{tab:fingerprint_comparison} summarizes the access assumptions and verification evidence of these methods.

Under white-box access, the verifier can control the generation pipeline or inspect intermediate model information.
FingerInv~\cite{teng2025fingerprinting} constructs optimized latent-noise fingerprints through DDPM inversion and requires a controllable DDPM-compatible denoising process with access to intermediate denoising information.
Its final ownership decision is based on whether a predefined QR-code pattern can be recovered from the generated image.
DiffIP~\cite{li2025diffip} extracts representation fingerprints from internal denoising features and verifies ownership by comparing the representations of the source and suspect models.
The two methods therefore rely on different ownership evidence despite both operating under white-box access.

Under black-box access, the verifier can only submit text queries and observe the final generated images.
TVN~\cite{guo2025one} realizes model fingerprints as text prompts with optimized non-transferable suffixes and verifies ownership from target-specific semantic deviations in the resulting images.
Although it supports end-to-end API verification, its optimized suffixes may exhibit linguistic abnormality. Moreover, since the generated outputs can deviate from the semantics of ordinary user queries, the verification queries may be more easily detected or filtered.

Built on collapsed generation, our framework supports both white-box and black-box ownership verification using the same behavioral evidence: statistically abnormal cross-seed consistency in the final generated images.
For white-box verification, although pipeline control is required to apply optimized continuous conditions, the ownership decision relies only on the final outputs rather than intermediate denoising states or internal representations, avoiding the internal-information dependence of existing white-box methods.
For black-box verification, our framework uses naturally collapsed prompts instead of optimized adversarial queries, making the verification queries closer to ordinary service interactions and less distinguishable by query detection or filtering mechanisms.

\subsection{Collapsed Generation in Diffusion Models}

Prior studies have shown that diffusion models can memorize and reproduce training examples under particular input conditions~\cite{carlini2023extracting,somepalli2023diffusiona,somepalli2023understanding,webster2023reproducible}.
Carlini et al.~\cite{carlini2023extracting} demonstrated that text-to-image diffusion models can regenerate individual training examples with high fidelity, enabling large-scale training data extraction.
Subsequent studies investigated the factors underlying such behavior, including data duplication~\cite{somepalli2023understanding}, the geometry of the learned probability distribution~\cite{jeon2025understanding,ross2025geometric}, training dynamics~\cite{wen2024detecting}, and localized model parameters or subspaces~\cite{chavhan2024memorization,hintersdorf2024finding}.

To identify memorized content, existing extraction methods search for prompts that trigger repeatable generations and compare the resulting images with reference training samples using pixel- or feature-level similarity measures~\cite{carlini2023extracting,somepalli2023diffusiona,pizzi2022self}.
Together, these findings suggest that particular input conditions may induce highly concentrated generation behavior, under which independent stochastic samplings yield visually similar outputs.
We refer to this phenomenon as \emph{collapsed generation}.
Its dependence on the learned data distribution, training dynamics, and model parameters motivates our investigation of collapsed generation as a model-dependent behavioral signature for ownership verification.

%% file: tables/fingerprint_comparison.tex
\begin{table*}[t]
    \centering
    \caption{Methodological comparison of non-invasive diffusion model fingerprints. White-box and black-box refer to the verifier's access to the suspect model. The table compares the input conditions and evidence used for verification, while further specifying the model control or observations required by each method.}
    \label{tab:fingerprint_comparison}
    \small
    \renewcommand{\arraystretch}{1.15}
    \setlength{\tabcolsep}{4pt}
    \begin{tabularx}{\textwidth}{@{}l l >{\raggedright\arraybackslash}p{0.17\textwidth} X >{\raggedright\arraybackslash}p{0.24\textwidth} c@{}}
        \toprule
        \textbf{Method} & \textbf{Verification Access} & \textbf{Input Condition for Verification} & \textbf{Verification Evidence} & \textbf{Required Control or Observation} & \textbf{Natural Query} \\
        \midrule
        FingerInv~\cite{teng2025fingerprinting} & White-box & Optimized latent noise & QR-code recovery from the final image & DDPM-compatible denoising control and intermediate denoising outputs & N/A \\
        DiffIP~\cite{li2025diffip} & White-box & Inputs for representation extraction & Representation-fingerprint similarity & Internal denoising representations of both models & N/A \\
        \textbf{Ours} & White-box & Optimized continuous prompt embedding & Cross-seed output consistency & Text-embedding injection and final-image observation & N/A \\
        \midrule
        TVN~\cite{guo2025one} & Black-box & Text prompt with an optimized adversarial suffix & Target-specific semantic deviation & Text-query API and final-image observation & No \\
        \textbf{Ours} & Black-box & Natural collapsed prompt & Cross-seed output consistency & Text-query API and final-image observation & Yes \\
        \bottomrule
    \end{tabularx}
\end{table*}

%% file: tex/prelim.tex
\section{Preliminaries}

\subsection{Threat Model}
We consider an IP protection scenario in which a model owner develops a proprietary T2I diffusion model, referred to as the \emph{source model}.
An adversary may obtain the source model without authorization and redistribute it directly, produce a modified or obfuscated derivative, or deploy it as a commercial service.
The model under examination is referred to as the \emph{suspect model}.
The objective of the owner, or an authorized verifier acting on the owner's behalf, is to determine whether the suspect model originates from or is derived from the source model.

A complete model fingerprinting framework involves both fingerprint construction and ownership verification.
During fingerprint construction, we assume that the owner has full access to the source model and its training process.

To address potential ownership ambiguity arising from competing fingerprint claims, we assume a standard commitment-based verification protocol~\cite{adi2018turningweaknessstrengthwatermarking}.
Before deployment, the owner commits the fingerprint set to a trusted third party.
A valid commitment with a timestamp preceding competing claims establishes the priority of the owner's fingerprint evidence.

For ownership verification, we consider two settings that differ in the verifier's access to the suspect model: 

\noindent
\textbf{White-box verification.}
In this setting, the suspect model is available as a checkpoint, for example, after unauthorized open-source redistribution~\cite{zeng2024huref} or when it is lawfully obtained for forensic examination~\cite{Uchida_2017,rouhani2018deepsignsgenericwatermarkingframework}.
The verifier has access to the model parameters and architecture and can observe and control the inference pipeline, including its latent variables, conditioning inputs, and intermediate denoising states.

\noindent
\textbf{Black-box verification.}
In this setting, the suspect model is deployed as an end-to-end T2I API or online service~\cite{fernandez2023stable,min2024watermarkconditioneddiffusionmodelip,adi2018turningweaknessstrengthwatermarking}.
The verifier can submit text prompts and observe the corresponding generated images, but has no access to the model parameters, architecture or intermediate activations.

We assume that the adversary may apply model-level obfuscations, such as fine-tuning, pruning, and quantization, or, in the black-box setting, query-time interventions to evade verification, while preserving acceptable generation quality for normal use.

\subsection{Principles of Model Fingerprinting}  \label{sec:prelim}
Following prior work~\cite{teng2025fingerprinting}, we consider three desirable properties for diffusion model fingerprints:

\noindent
\textbf{Uniqueness.}
The fingerprint should be sufficiently distinctive to uniquely identify the target model among others.  
In particular, a fingerprint constructed from one source model should not be validated on an unrelated suspect model.

\noindent
\textbf{Robustness.}  
The fingerprint should remain verifiable even after the model undergoes obfuscations.  
This property prevents adversaries from invalidating the verification.

\noindent
\textbf{Stealthiness.}  
For black-box verification, fingerprint queries should appear natural and resemble ordinary service interactions.
This reduces the likelihood that the queries are detected or filtered by the suspect service.

\subsection{Diffusion Models}
Diffusion models generate data by iteratively denoising a random noise sample through a learned reverse process~\cite{ho2020denoisingdiffusionprobabilisticmodels}. 
Let $\hat{x}_0$ denote the clean image, and $\{x_t\}_{t=1}^T$ represent its progressively noised versions obtained via a forward diffusion process controlled by a variance schedule $\{\beta_t\}$:
\begin{equation}
x_t = \sqrt{\bar{\alpha}_t}\hat{x}_0 + \sqrt{1-\bar{\alpha}_t}\epsilon, \quad \epsilon \sim \mathcal{N}(0, I),
\end{equation}
where $\bar{\alpha}_t = \prod_{i=1}^t (1 - \beta_i)$.

The reverse denoising process is parameterized by a neural network $\epsilon_\theta$ that predicts the noise added at each step. 
Given a noisy sample $x_t$, a single reverse step reconstructs $x_{t-1}$ as:
\begin{equation}
x_{t-1} = \frac{1}{\sqrt{\alpha_t}}\left(x_t - \frac{1-\alpha_t}{\sqrt{1-\bar{\alpha}_t}}\,\epsilon_\theta(x_t, t)\right) + \sigma_t z_t, 
\end{equation}
where $\sigma_t$ is the step-dependent noise scale and $z_t \sim \mathcal{N}(0, I)$ is random noise. 

In T2I diffusion models~\cite{rombach2022high}, additional condition $c$ is derived from a text encoder $E_{\text{text}}$ that maps a prompt $p$ to an embedding $c = E_{\text{text}}(p)$. 
This augments the denoiser $\epsilon_\theta$ with text embedding $c$, enabling conditional generation from a prompt $p$. 
The overall sampling process from noise $x_T \sim \mathcal{N}(0, I)$ to the generated image $x_0$ can be expressed as a function of the prompt:
\begin{equation}
x_0 = G_\theta(x_T, c = E_{\text{text}}(p)),
\end{equation}
where $x_T$ controls the stochasticity of the generation. 

For a given prompt $p$, different initial samples $x_T$ typically yield diverse outputs, whereas consistent and low-diversity outputs indicate a \emph{collapsed generation} behavior.
This behavior forms the foundation for analyzing model-specific generative characteristics, which are exploited in our fingerprinting framework.

%% file: tex/method.tex
\input{figs/density}

\input{figs/manifold}

\section{Methodology}

\subsection{Collapsed Generation in Diffusion Models}
Recent studies have shown that diffusion models can memorize and reproduce a subset of their training data~\cite{carlini2023extracting,somepalli2023understanding}.
One manifestation of such memorization is reduced generation diversity. For certain prompts associated with memorized samples, independent stochastic samplings can produce highly similar outputs~\cite{ross2025geometric,jeon2025understanding}.
This behavior contrasts with typical diffusion generation, where different initial noises are expected to yield diverse images under the same prompt.
Although diffusion models are generally less prone to mode collapse than GANs~\cite{zhang2018convergence,miao2024training}, this prompt-specific concentration exhibits a similar loss of output diversity.
We term this phenomenon \emph{collapsed generation} and formalize it through cross-seed consistency.

\begin{definition}[Collapsed Generation]
\label{def:collapsed_generation}
Let $c$ denote the conditioning embedding supplied to a diffusion model $G_\theta$.
For a text prompt $p$, $c=E_{\text{text}}(p)$.
Given $K$ independently sampled initial noises
$\{x_T^{(k)}\}_{k=1}^K$, let
\[
x_0^{(k)}(c)=G_\theta(x_T^{(k)},c)
\]
denote the corresponding generated images.
Their cross-seed consistency is measured as
\begin{equation}
\label{eq:similarity}
\bar{s}_\theta(c,K)
=
\frac{2}{K(K-1)}
\sum_{i=1}^{K-1}
\sum_{j=i+1}^{K}
s\!\left(x_0^{(i)}(c),x_0^{(j)}(c)\right),
\end{equation}
where $s(\cdot,\cdot)$ is a perceptual similarity metric.
The generation conditioned on $c$ is considered \emph{collapsed} if
\begin{equation}
\bar{s}_\theta(c,K)\geq\tau_s,
\end{equation}
where $\tau_s$ is the collapse detection threshold.
\end{definition}

When the model is clear from context, we omit the subscript $\theta$.
For a text prompt $p$, we use $\bar{s}(p,K)$ as shorthand for
$\bar{s}_\theta(E_{\text{text}}(p),K)$.

In this work, we instantiate $s(\cdot,\cdot)$ using the Self-Supervised Copy Detection (SSCD) score~\cite{pizzi2022self}.
SSCD computes the cosine similarity between image representations extracted by a self-supervised model trained for image copy detection.
Its sensitivity to visual and structural correspondence makes it suitable for measuring output consistency.

Collapsed generation represents an atypical yet repeatable behavior of the learned generation process.
As shown in Fig.~\ref{fig:density}(a), prompts that induce collapse on SD~1.4 form a high-consistency distribution that is clearly separated from that of normal prompts.
This separation makes the model's intrinsic collapse behavior observable through cross-seed similarity, suggesting its potential as a behavioral fingerprint.
We next investigate whether such collapse patterns are specific to the model that produces them.

\input{figs/overview}

\subsection{Model Fingerprint via Collapsed Generation}

For collapsed generation to serve as a model fingerprint, the collapse-inducing condition should be specific to the learned generation process.
Given a source model, a fingerprint condition is expected to produce high cross-seed consistency on the source model while eliciting substantially weaker consistency on independently trained models.
This property corresponds to the \emph{uniqueness} requirement of model fingerprinting.

We first examine this property using naturally occurring collapsed prompts from SD~1.4 and SD~2.1.
Specifically, we apply the collapsed prompts identified for each source model to both models and measure their cross-seed consistency.
As shown in Fig.~\ref{fig:density} (b), matched prompt--model pairs produce substantially higher consistency scores than mismatched pairs.
The same prompt therefore induces different collapse responses across models, showing that collapsed generation reflects characteristics of the learned model rather than the input condition alone.

This model dependence arises from differences in the learned conditional generation mappings.
Training data, model architecture, and optimization dynamics jointly shape how a conditioning input is mapped to the generative manifold~\cite{somepalli2023understanding,somepalli2023diffusiona,ross2025geometric}.
Consequently, a condition located in a collapsed region of one model may produce diverse generations in another, as conceptualized in Fig.~\ref{fig:manifold}.
These model-dependent collapsed regions provide the behavioral basis for constructing model fingerprints.

The comparison above involves T2I models that may differ in their training data or model configurations.
For ownership verification, a stronger requirement is that collapsed fingerprints distinguish independently trained model instances even when these factors are controlled.
We examine this requirement in Sec.~\ref{sec:cifar_exp}, where the model architecture, training data, and optimization procedure are held fixed, while only the parameter initialization is varied.
The resulting models exhibit distinguishable collapsed fingerprints, showing that the learned collapse behavior can be specific to an individual training outcome.

This result is also relevant to models trained from shared public resources.
Even when independently trained models use the same public dataset, different parameter initializations can lead to different learned parameters and memorization patterns, thereby producing distinct collapsed regions~\cite{carlini2023extracting,somepalli2023understanding}.
The same mechanism provides a basis for differentiating modern large-scale T2I models, while practical differences in data curation, architecture, and training pipelines introduce further variation into their learned generation behavior.
Such training-dependent collapse behavior provides the granularity required for ownership verification.
Having established the observability and model specificity of collapsed generation, we next describe how the corresponding fingerprint conditions are constructed.

\subsection{Fingerprint Construction} \label{sec:fp_construct} 
Having established the model specificity of collapsed generation, the next challenge is to turn this behavioral signal into reproducible fingerprint conditions.
The inputs that induce collapsed generations are not known a priori, and their admissible form is determined by the interface available during verification.
With access to the inference pipeline, the continuous conditioning space can be actively searched to synthesize fingerprint embeddings. Under end-to-end T2I API-only access, collapse must instead be elicited by text prompts.
We address these two settings through pipeline-access continuous embedding synthesis and API-only natural-prompt mining, respectively.

\input{figs/visual_mem_vs_nonmem}

\noindent
\textbf{Continuous-Embedding Synthesis.}
In checkpoint-level ownership verification, white-box access gives the verifier control over the suspect model's inference pipeline, allowing continuous conditioning embeddings to be supplied directly.
This makes continuous embeddings a viable representation for fingerprint conditions.
The owner therefore constructs such fingerprints by optimizing embeddings on the source model to induce collapsed generation, as illustrated in Fig.~\ref{fig:overview} (a).

We construct multiple embedding fingerprints for joint evaluation during verification.
For the $i$-th fingerprint, we initialize
$c_i=E_{\mathrm{text}}(p_i)$
and sample $K$ independent initial noises
$\{x_T^{(k)}\}_{k=1}^{K}$.
We then optimize $c_i$ such that the corresponding sampling trajectories suppress the variation introduced by different initial noises and converge toward similar outputs.

A direct approach would maximize the final-image consistency $\bar{s}_\theta(c_i,K)$ defined in Eq.~\ref{eq:similarity}.
However, each optimization iteration would require differentiating through the complete denoising process, decoding all resulting latents into images, and evaluating their perceptual similarity.
This makes direct optimization computationally and memory intensive, particularly when multiple trajectories are jointly optimized~\cite{chen2023content,lin2023sd}.

In this work, our key observation is that the characteristic output content of collapsed generation becomes established early in the denoising process.
As illustrated in Fig.~\ref{fig:mem_vs_nonmem_steps}, for a collapsed example, the predicted clean image rapidly develops a stable and recognizable structure within the first few denoising steps.
In contrast, for a normal example, the prediction remains ambiguous at the early stage and becomes progressively clearer as denoising proceeds.
This contrast suggests that the generation tendency underlying collapse is already reflected in early intermediate states.
Motivated by this observation, we adopt \emph{truncated optimization}, which minimizes the dispersion among early intermediate latents from independent initial noises as a surrogate for final-image consistency.

Let $x_{T-t_{\mathrm{trunc}}}^{(k)}(c_i)$ denote the intermediate latent obtained from $x_T^{(k)}$ after the first $t_{\mathrm{trunc}}$ denoising steps, and let
\begin{equation}
\bar{x}_{T-t_{\mathrm{trunc}}}(c_i)
=
\frac{1}{K}
\sum_{k=1}^{K}
x_{T-t_{\mathrm{trunc}}}^{(k)}(c_i)
\end{equation}
denote their mean.
We define the collapse loss as
\begin{equation}
\label{eq:collapse_loss}
\mathcal{L}_{\mathrm{Collapse}}(c_i)
=
\frac{1}{K}
\sum_{k=1}^{K}
\left\|
x_{T-t_{\mathrm{trunc}}}^{(k)}(c_i)
-
\bar{x}_{T-t_{\mathrm{trunc}}}(c_i)
\right\|_2^2.
\end{equation}
Minimizing this objective drives the early denoising trajectories toward a shared latent region.
Since gradients are propagated through only the first $t_{\mathrm{trunc}}$ steps, the remaining denoising process, VAE decoding, and image-level similarity computation are excluded from the optimization loop, thereby reducing both computational and memory costs.

Repeating this procedure from $M$ initial embeddings produces the fingerprint set
\begin{equation}
\mathcal{C}_f
=
\left\{
c_i^*
\mid
c_i^* = \arg\min_{c_i}
\mathcal{L}_{\mathrm{Collapse}}(c_i),
\ i=1,\ldots,M
\right\}.
\end{equation}
After optimization, the owner evaluates each $c_i^*$ through complete sampling on the source model and measures its final-image cross-seed consistency using Eq.~\ref{eq:similarity}.
The detailed optimization procedure is provided in Appendix~\ref{sec:apdx_algorithm}.

\input{figs/training_loss}

\noindent
\textbf{Natural-Prompt Mining.}
When the suspect model is exposed only through an end-to-end T2I API, the continuous embeddings constructed above cannot be directly replayed.
In black-box verification, fingerprint conditions must therefore be represented as text prompts.
We further construct them from naturally occurring prompts associated with the source model's training data, allowing the resulting fingerprint queries to retain the form of ordinary API inputs.

The main challenge is to efficiently identify prompts that induce collapsed generation among the large-scale training data of modern T2I models.
A direct search would require generating $K$ images for every candidate prompt and evaluating their cross-seed consistency.
Applying this procedure to a large training collection would incur a substantial number of complete sampling runs.

We address this search cost by exploiting the relationship between collapsed generation and sample-level training loss~\cite{bonnaire2025why}.
As shown in Fig.~\ref{fig:training_loss}, training samples associated with collapsed generation are substantially enriched in the low-loss region.
The model owner can therefore record per-sample denoising losses during training and retain the prompts associated with low-loss samples as a candidate pool $\mathcal{P}_{\mathrm{cand}}$, as illustrated in Fig.~\ref{fig:overview} (b).

Only this reduced candidate pool requires output-level evaluation.
For each $p\in\mathcal{P}_{\mathrm{cand}}$, the owner generates $K$ samples on the source model and computes the cross-seed consistency $\bar{s}(p,K)$.
The owner selects the $M$ prompts with the highest consistency scores to form
$\mathcal{P}_f=\{p_i^*\}_{i=1}^{M}$.
The resulting prompts can subsequently be submitted to the suspect model through its standard text interface.

The two construction strategies yield interface-compatible fingerprint sets $\mathcal{C}_f$ and $\mathcal{P}_f$, both of which are evaluated through the cross-seed consistency test described next.

\subsection{Fingerprint Verification}
\label{sec:fp_verify}

Observing a high consistency score on the suspect model is not by itself sufficient for ownership verification.
Normal prompts may occasionally produce similar outputs, while an individual fingerprint condition may exhibit an atypical response.
The main challenge is therefore to determine whether the collapse induced by the fingerprint set is statistically distinguishable from the source model's normal generation behavior.
We address this challenge through source-referenced statistical calibration and set-level aggregation.

The same verification principle applies to both embedding and text fingerprints.
Let $G_{\theta'}$ denote the suspect model.
For an embedding fingerprint $c_m^*\in\mathcal{C}_f$, the verifier supplies the embedding directly to the suspect pipeline; for a natural-prompt fingerprint $p_m^*\in\mathcal{P}_f$, the verifier submits the corresponding text prompt.
In either case, $K$ independent outputs are collected and their cross-seed consistency is computed using Eq.~\ref{eq:similarity}.
The responses of the $M$ fingerprints are then aggregated as
\begin{equation}
\label{eq:fingerprint_statistic}
\bar{s}_f
=
\frac{1}{M}
\sum_{m=1}^{M}
\bar{s}_{\theta'}(c_m^*,K)
\end{equation}
for $\mathcal{C}_f$, or equivalently
$\bar{s}_f=\frac{1}{M}\sum_{m=1}^{M}\bar{s}_{\theta'}(p_m^*,K)$
for $\mathcal{P}_f$.

Before deployment, the owner establishes a normal-generation reference sample on the source model using a set of normal prompts, which are generated independently by GPT-4 in our implementation.
Let
$\{u_n=\bar{s}_\theta(p_n,K)\}_{n=1}^{N}$
denote the resulting reference scores, and let $\mu_0$ and $\sigma_0$ be their sample mean and standard deviation.
Treating the aggregate fingerprint response $\bar{s}_f$ as a new observation relative to this reference sample, we perform a right-tailed predictive $t$-test:
\begin{equation}
\label{eq:fingerprint_test}
t_f
=
\frac{
\bar{s}_f-\mu_0
}{
\sigma_0\sqrt{1+1/N}
},
\qquad
p_{\mathrm{val}}
=
1-F_{t_{N-1}}(t_f),
\end{equation}
where $F_{t_\nu}$ denotes the Student's $t$ CDF with $\nu$ degrees of freedom.

A small $p_{\mathrm{val}}$ indicates that the aggregate fingerprint response lies in the upper tail of the source model's normal-generation reference, providing evidence that the suspect model reproduces the source model's collapsed fingerprint behavior.
Given a decision threshold $\tau_f$, we accept the ownership claim when
\begin{equation}
p_{\mathrm{val}}<\tau_f.
\end{equation}
The model specificity established above makes this decision selective across models: independently trained models are not expected to reproduce the same set-level response, providing the \textit{uniqueness} required for ownership verification.
Aggregating multiple fingerprint responses further reduces the influence of an anomalous individual condition.
Moreover, the normal-generation reference can be prepared before deployment and reused without requiring additional normal-prompt queries to the suspect model.
The detailed procedure is provided in Appendix~\ref{sec:apdx_algorithm}.

For API-only verification, an additional practical challenge is to prevent the fingerprint queries from being readily distinguished from normal service interactions, i.e., \textit{stealthiness}.
Our natural-prompt fingerprints preserve the linguistic form of normal user inputs, avoiding the conspicuous token patterns introduced by adversarially optimized prompts, as evaluated in Sec.~\ref{sec:black_exp}.
Beyond the appearance of individual prompts, stealthiness also depends on the pattern in which repeated queries are issued.
The verification statistic requires $K$ independent generations for each prompt, but these generations need not be obtained consecutively or within a single session.
They may be collected over temporally separated interactions and, when appropriate, pooled across multiple verification clients before applying the same statistical test.
Under the default configuration of $M=4$ and $K=4$, this amounts to a total of $16$ fingerprint queries.
Together, the natural prompt form, modest query budget, and flexible collection schedule jointly support stealthy verification.
We further evaluate robustness against query-time interventions in Sec.~\ref{sec:black_exp}.

%% file: figs/density.tex
\begin{figure*}[t]
    \centering
    \begin{subfigure}[t]{0.33\linewidth} 
        \centering
        \includegraphics[height=4cm, keepaspectratio,width=0.96\linewidth]{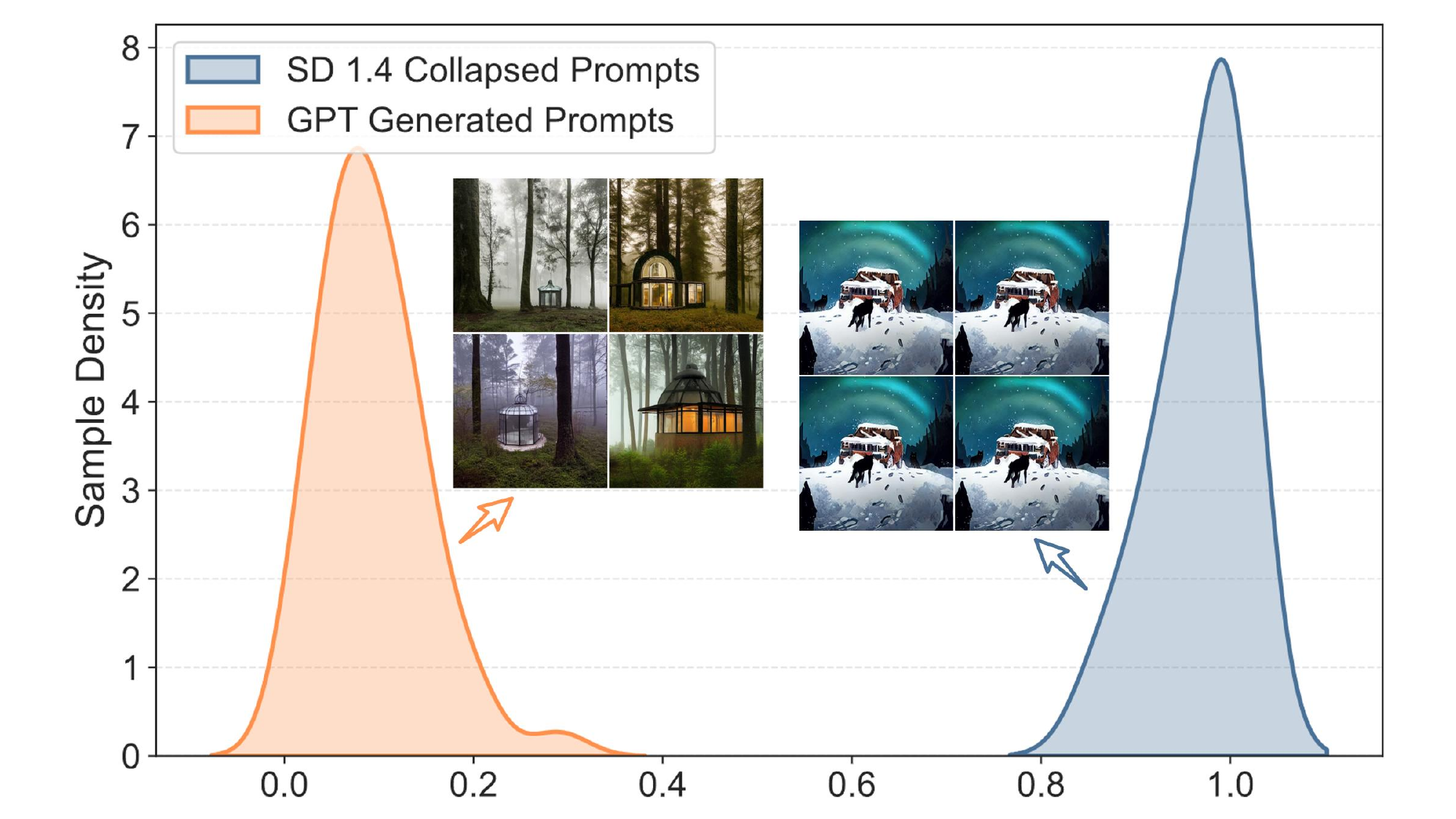}
        \caption{}
    \end{subfigure}
    \hfill
    \begin{subfigure}[t]{0.65\linewidth} 
        \centering
        \includegraphics[height=4cm, keepaspectratio,width=0.48\linewidth]{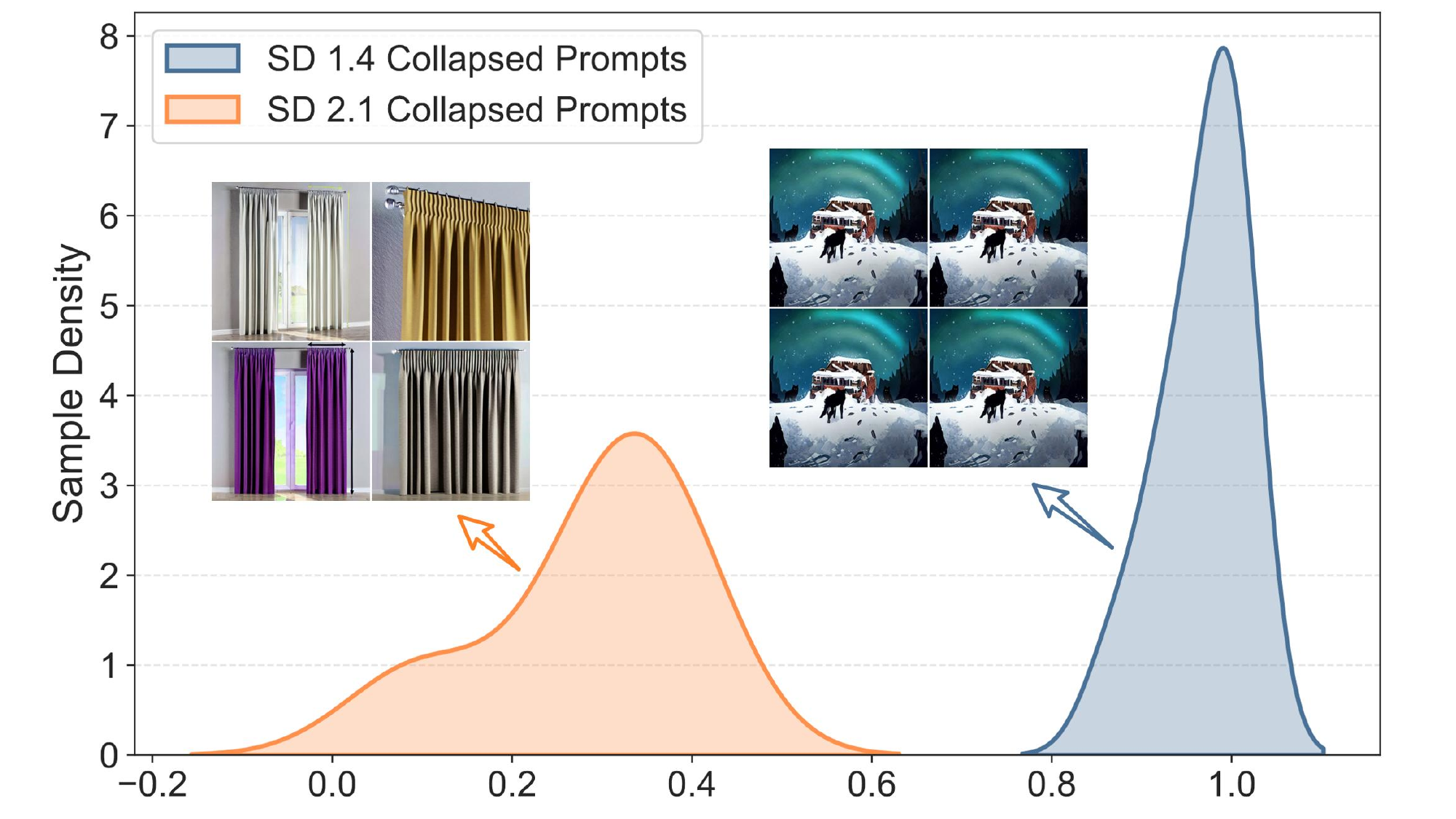}
        \hfill
        \includegraphics[height=4cm, keepaspectratio,width=0.48\linewidth]{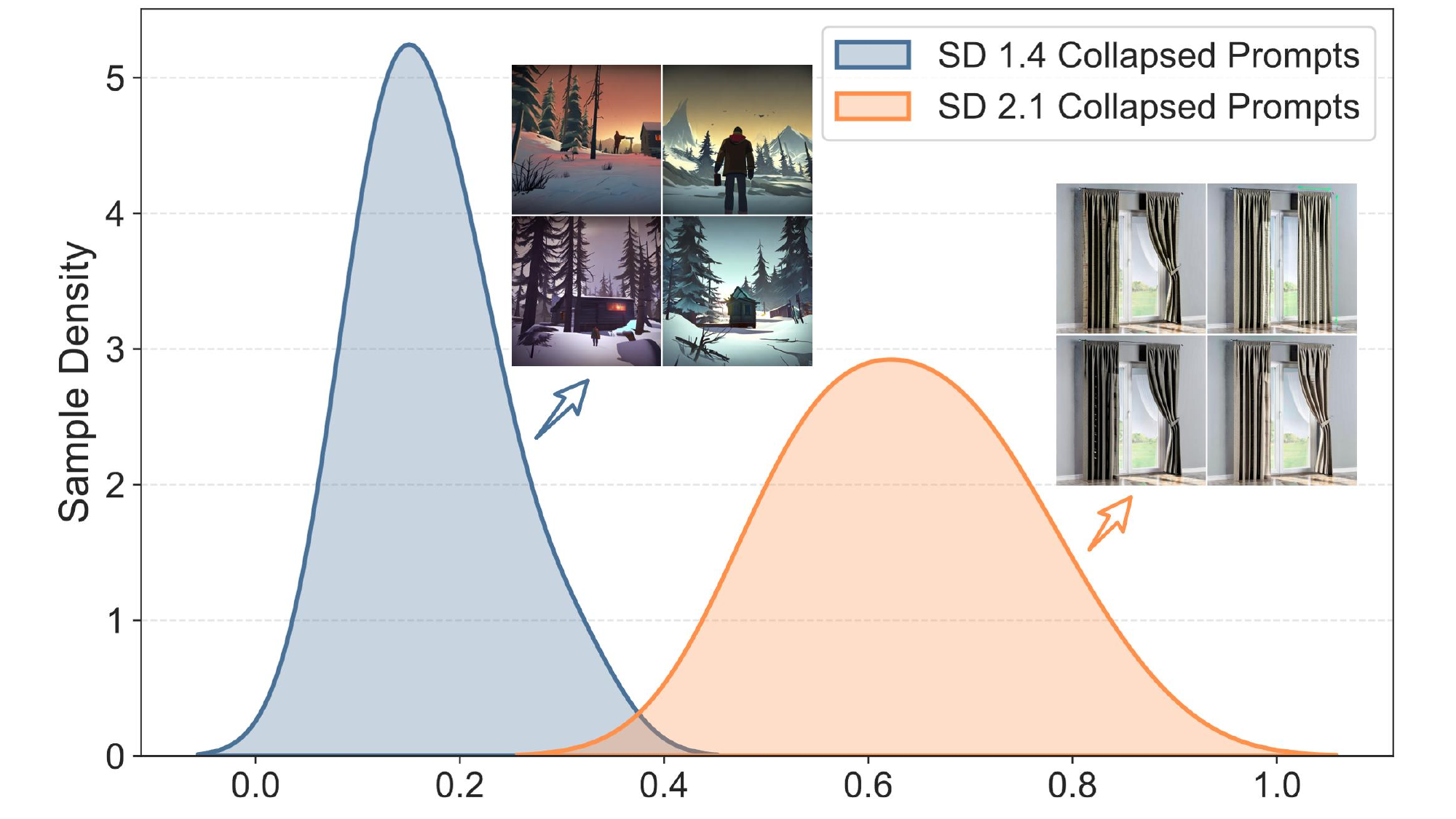}
        \caption{}
    \end{subfigure}
    \caption{
        (a): Distribution of average pairwise similarity scores $\bar{s}(p, K=4)$ for normal prompts vs. collapsed prompts on SD 1.4. 
        (b): Distribution of $\bar{s}(p, K=4)$ for SD 1.4 collapsed prompts vs. SD 2.1 collapsed prompts on SD 1.4 (left) and SD 2.1 (right), respectively.
        Collapsed generations not only differ from normal ones but also exhibit model-specific characteristics, enabling effective fingerprinting.
    }
    \label{fig:density}
\end{figure*}

%% file: figs/manifold.tex
\begin{figure}
    \centering
    \includegraphics[width=0.9\linewidth]{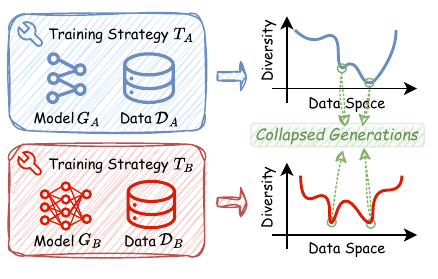}
    \caption{Different model architectures, training data, and optimization dynamics lead to distinct generative manifolds, resulting in model-specific collapsed generation behaviors.}
    \label{fig:manifold}
\end{figure}

%% file: figs/overview.tex
\begin{figure*}[t]
    \centering
    \includegraphics[width=0.9\linewidth]{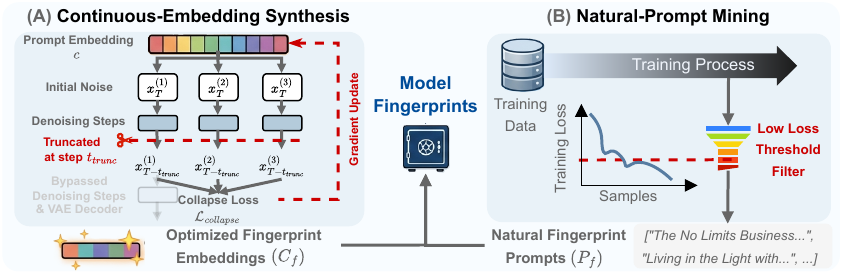}

    \caption{\textbf{Overview of fingerprint construction.}
    \textbf{(a) Continuous-Embedding Synthesis.}
    For pipeline-access verification, the owner optimizes conditioning embeddings on the source model.
    Starting from an initial embedding $c_i$ and $K$ independent noises $\{x_T^{(k)}\}_{k=1}^{K}$, truncated optimization minimizes the dispersion among early intermediate latents to produce $c_i^*$.
    Repeating this procedure yields the embedding fingerprint set $\mathcal{C}_f$.
    \textbf{(b) Natural-Prompt Mining.}
    For API-only verification, per-sample denoising loss is used to identify a low-loss candidate pool $\mathcal{P}_{\mathrm{cand}}$.
    The candidates are then evaluated by their cross-seed consistency on the source model, and the strongest prompts form the natural-prompt fingerprint set $\mathcal{P}_f$.}
    \label{fig:overview}
\end{figure*}

%% file: figs/visual_mem_vs_nonmem.tex
\begin{figure}[t]
    \centering
    \includegraphics[width=\linewidth]{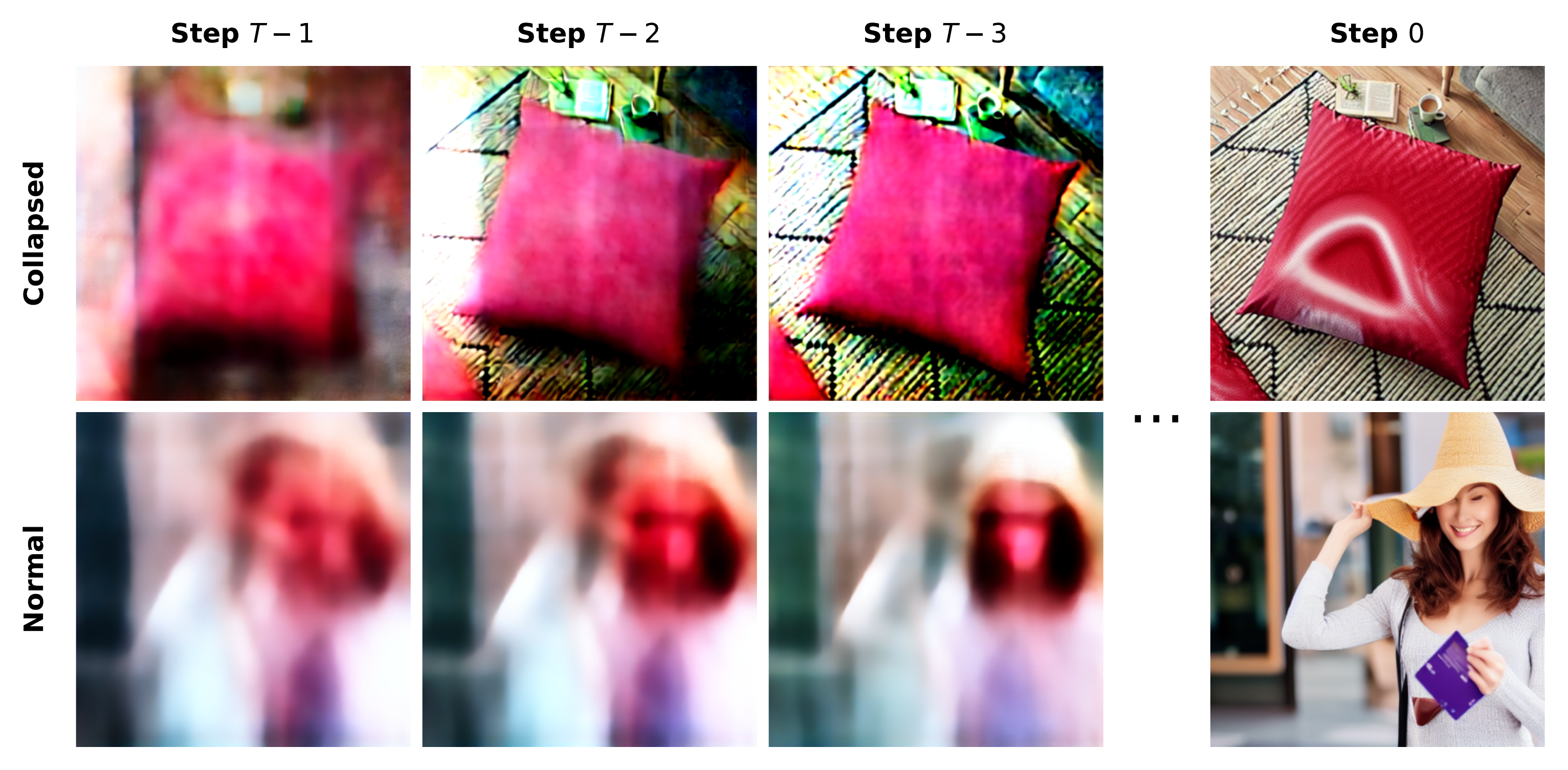}
    \caption{Visualization of predicted $x_0$ along the denoising process for a representative collapsed and normal example in SD~1.4.
    For the collapsed example (top), stable and recognizable output content emerges within the first few denoising steps.
    For the normal example (bottom), the prediction remains ambiguous at the early stage and becomes progressively clearer as denoising proceeds.}
    \label{fig:mem_vs_nonmem_steps}
\end{figure}

%% file: figs/training_loss.tex
\begin{figure}[t]
    \centering
    \includegraphics[width=0.88\linewidth]{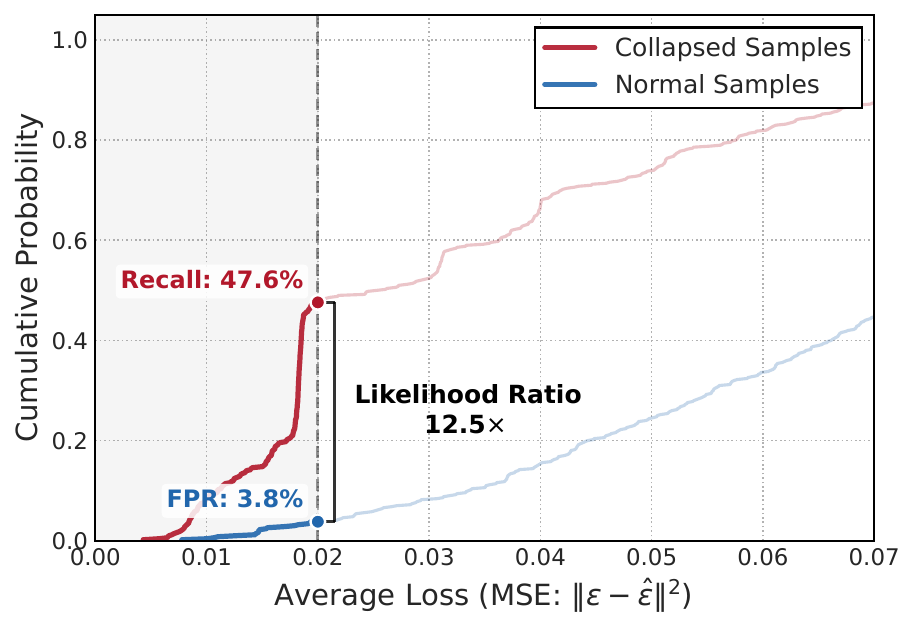}
    \caption{Cumulative distribution of training loss for collapsed and normal samples in SD 1.4. The loss is computed as the MSE between the ground-truth noise $\epsilon$ and the predicted noise $\hat{\epsilon}$ at intermediate timesteps ($t\in[450, 550]$)~\cite{zhai2024membership}. Collapsed samples are predominantly concentrated in the low-loss region. Consequently, applying a low-loss threshold yields a significantly higher proportion of collapsed samples (e.g., a likelihood ratio of roughly $12.5$ at a threshold of $0.02$), enabling efficient extraction of fingerprint prompts.}
    \label{fig:training_loss}
\end{figure}

%% file: tex/experiment.tex
\section{Experiments}

Following the principles in Sec.~\ref{sec:prelim}, we organize our evaluation around four research questions.

\noindent
\textbf{RQ1: Uniqueness.} Are collapsed-generation fingerprints specific to their source models and distinguishable from independently trained models?

\noindent
\textbf{RQ2: Robustness.} Do the fingerprints remain verifiable after model modifications and adaptive query-time obfuscations?

\noindent
\textbf{RQ3: Stealthiness.} In the API-only setting, do fingerprint queries resemble ordinary user prompts and remain difficult to detect?

\noindent
\textbf{RQ4: Efficiency.} Can collapsed-generation fingerprints be constructed with practical computational overhead?

In the following, we first present a controlled preliminary study using conditional DDPMs to validate the uniqueness of collapsed-generation fingerprints.
We then describe the common experimental setup for T2I models and evaluate fingerprint uniqueness and robustness under both white-box pipeline access and black-box API-only access.
For the latter, we additionally assess the stealthiness of fingerprint queries.
Finally, we analyze the computational cost of fingerprint construction.

\subsection{Preliminary Study: Model-Specific Collapse}
\label{sec:cifar_exp}

\input{figs/cifar}

To answer RQ1, we conduct a controlled study using independently trained conditional DDPMs on CIFAR-10 to examine whether collapsed-generation fingerprints remain specific to individual training runs.

\noindent
\textbf{Experimental Settings.}
We construct a conditional DDPM for CIFAR-10 in which each training image is assigned an instance-specific condition composed of its class label and image index.
The class label and image index are encoded separately and then fused as the conditioning input to the U-Net.
This conditioning design allows collapsed generation to be measured at the level of individual training samples.

We independently train four model instances with the same architecture, dataset, optimizer, and training schedule, using different random seeds for parameter initialization and training stochasticity.
To accelerate the emergence of pronounced collapse cases, we select the same $500$ CIFAR-10 training images as an upweighted subset for all four models and sample them approximately $10\times$ more frequently than the remaining images.
The identical exposure schedule enables a controlled comparison of collapse patterns across independent training runs.
Each model is trained for $40{,}000$ steps with a batch size of $128$ and evaluated using the final checkpoint, the same DDPM sampler, a classifier-free guidance scale of $1.5$, and four fixed sampling seeds shared across all models.

\noindent
\textbf{Fingerprint Construction and Verification.}
Following the low-loss mining strategy described in Sec.~\ref{sec:fp_construct}, we rank the training set for each model according to sample-level denoising loss during training and retain the low-loss conditions as fingerprint candidates.
For each candidate, we generate four images and measure cross-seed diversity using the average pairwise $\ell_2$ distance among the generated images, where a smaller distance indicates stronger collapsed generation.
The four candidates with the lowest distances form the fingerprint set of each model.

To establish the normal-generation reference, we randomly sample $50$ non-upweighted training conditions for each model and compute their cross-seed $\ell_2$ distances.
Following the set-level verification procedure in Sec.~\ref{sec:fp_verify}, we evaluate each four-condition fingerprint set on all four models and obtain a $4\times4$ matrix of lower-tail $p_{\mathrm{val}}$ scores.

\noindent
\textbf{Results.}
Fig.~\ref{fig:cifar} (a) exhibits a clear diagonal separation, with matched fingerprint--model pairs yielding markedly lower $p_{\mathrm{val}}$ scores than all mismatched pairs.
This result indicates that the collapse-prone conditions identified from one training run are not reproduced with comparable strength by the other independently trained models.
The qualitative examples in Fig.~\ref{fig:cifar} (b) corroborate this distinction: each representative fingerprint produces consistent cross-seed generations on its source model, whereas the fingerprint selected from the first source model yields substantially more diverse outputs on the three mismatched models.
Together, these results show that model-specific collapse patterns emerge across independent training runs under matched architectures, training data, upweighted samples, and optimization configurations, supporting the uniqueness of collapsed-generation fingerprints.

\subsection{Experimental Setup}

Having established model-specific collapse in the controlled CIFAR-10 setting, we next evaluate the proposed fingerprinting framework on representative T2I diffusion models.
The following experiments use SSCD-based cross-seed similarity and the T2I fingerprint construction and verification procedures described in Secs.~\ref{sec:fp_construct} and~\ref{sec:fp_verify}. 
We describe the experimental setup below.

\noindent
\textbf{Models.}
We consider both white-box pipeline-access and black-box API-only verification settings.
Under pipeline access, we evaluate Stable Diffusion 1.4 (SD 1.4)~\cite{rombach2022high}, DeciDiffusion v1.0 (Deci)~\cite{DeciFoundationModels}, PixArt-$\alpha$ (PixArt)~\cite{chen2024pixartalpha}, and Stable Diffusion 3 (SD 3)~\cite{esser2024scaling}.
These models span both U-Net-based and diffusion Transformer (DiT)-based architectures, with PixArt and SD 3 representing the latter.
In particular, SD 3 combines a flow-matching formulation with training-data deduplication, providing a challenging case for assessing whether collapsed-generation fingerprints remain effective when naturally occurring collapse may be less prevalent.
Under API-only access, we evaluate SD 1.4 and Stable Diffusion 2.1 (SD 2.1)~\cite{stabilityai_sd21_2022}, for which prior work provides public prompt collections that support reproducible candidate construction.
All pretrained checkpoints are obtained from Hugging Face~\cite{hugging}, with their exact versions reported in Appendix~\ref{sec:apdx_exp_setting}.

\noindent
\textbf{Fingerprint Construction.}
For pipeline-access verification, we construct the fingerprint set $\mathcal{C}_f$ by optimizing continuous conditioning embeddings independently on each source model.
For API-only verification, we draw candidates from the public prompt collections released by~\cite{webster2023reproducible} for studying black-box training-data extraction.
Each collection contains $500$ prompts for which the corresponding model was shown to reproduce training examples.
We use these prompts as an enriched candidate pool for collapsed generation, screen them according to their cross-seed generation consistency, and select the top-$M$ prompts to form $\mathcal{P}_f$.
Unless otherwise specified, both $\mathcal{C}_f$ and $\mathcal{P}_f$ contain $M=4$ fingerprint conditions, and each condition is evaluated using $K=4$ generations.

\noindent
\textbf{Inference Configuration.}
For SD 1.4, SD 2.1, and Deci, we use DDIM sampling with $50$ denoising steps and a classifier-free guidance scale of $7.5$.
PixArt uses its default DPM-Solver scheduler with $20$ denoising steps, while SD 3 uses its default $28$-step sampler and a guidance scale of $7.0$.
Four fixed sampling seeds are shared across all applicable models and conditions.

\noindent
\textbf{Obfuscations.}
To evaluate robustness against model-level modifications, we consider pruning, quantization, and fine-tuning.
For pruning, we remove $10\%$, $20\%$, and $30\%$ of the smallest-magnitude weights from the model backbone, including the U-Net or Transformer blocks.
For quantization, we evaluate bfloat16, int8, and fp4 parameter representations.
For fine-tuning, we consider three representative derivatives of SD 1.4: Stable Diffusion 1.5 (SD 1.5)~\cite{rombach2022high}, Deliberate v4~\cite{Deliberate}, and Realistic Vision v2.0~\cite{RealisticVision}.
In the API-only setting, we additionally evaluate four adaptive query-time obfuscations: GPT rewriting, Random Token Addition (RTA)~\cite{somepalli2023understanding}, Embedding Optimization (EO)~\cite{wen2024detecting}, and Sharpness-Aware Initialization for Latent Diffusion (SAIL)~\cite{jeon2025understanding}, whose configurations are detailed in Appendix~\ref{sec:apdx_exp_setting}.

\noindent
\textbf{Baselines.}
Under pipeline access, we compare our method with FingerInv~\cite{teng2025fingerprinting}, which derives a model-specific latent code through DDPM inversion of a reference QR-code image and verifies model identity through its recovery on the queried model.
Under API-only access, we compare with TVN~\cite{guo2025one}, which searches for non-transferable adversarial prompt suffixes that induce target-specific semantic deviations.
Following its original configuration, TVN optimizes from $10$ initial prompts for each target model and retains the best-performing prompt as its fingerprint.
Detailed baseline implementations are provided in Appendix~\ref{sec:apdx_exp_setting}.

\noindent
\textbf{Metrics.}
For our method, we compute SSCD-based cross-seed generation consistency and derive the set-level $p_{\mathrm{val}}$ defined in Sec.~\ref{sec:fp_verify}.
For each source model, we construct the normal-generation reference using $N=50$ prompts independently generated by GPT-4.
We use $\tau_f=10^{-4}$ as the verification threshold throughout the T2I experiments, with $p_{\mathrm{val}}<\tau_f$ indicating a fingerprint match.
TVN measures the CLIP similarity between the generated image and its input prompt, whereas FingerInv measures the $\ell_2$ distance between the recovered and reference QR-code images.
For all three methods, lower reported values provide stronger evidence of a match under their respective verification rules.
For API-only stealthiness, we additionally measure prompt perplexity using Mistral-7B~\cite{jiang2023mistral7b}, where lower perplexity indicates a more natural prompt.
Additional optimization hyperparameters and computational resources are reported in Appendix~\ref{sec:apdx_exp_setting}.

\input{figs/wb_unique}

\subsection{White-box Pipeline-access Verification}

\input{figs/wb_robust_cases}

\noindent
\textbf{Optimization Details.} 
In the white-box setting, we induce collapsed generations by optimizing the prompt embedding. 
In the main experiments, we initialize the optimization from text embeddings of candidate prompts for SD v1 models. 
The initialization analysis in Appendix~\ref{sec:apdx_additional_results} further shows that both low-diversity and random prompt initializations yield verifiable fingerprints.
Further details and hyper-parameters are available in Appendix~\ref{sec:apdx_exp_setting}.

\noindent
\textbf{Uniqueness Analysis.}  
Fig.~\ref{fig:wb_unique} reports the cross-model verification results of our method and FingerInv under pipeline access.
Note that FingerInv is formulated upon DDPM inversion trajectories, making it incompatible with the underlying ODE mechanisms of flow-matching models like SD 3.

The results show that both our method and FingerInv exhibit excellent discriminative ability, with significantly low metric values (indicated by blue cells) appearing only along the main diagonal.
Cross-model verification is performed only between models with compatible conditioning interfaces.
Because PixArt and SD 3 use embedding spaces incompatible with the other evaluated pipelines, the corresponding cross-architecture entries are left undefined.
Under pipeline access, these interface incompatibilities can be identified before statistical fingerprint verification.

\noindent
\textbf{Robustness Analysis.}  
Both our method and FingerInv demonstrate strong verification performance under obfuscation attacks. 
The complete quantitative results are reported in Appendix~\ref{sec:apdx_additional_results}, while Fig.~\ref{fig:wb_robust_cases} provides representative qualitative examples.

As illustrated in the figure, FingerInv and our method exhibit different degradation behaviors.
FingerInv relies on recovering a specific reference pattern, and its generated QR codes exhibit visible degradation as model obfuscation becomes stronger.
In contrast, our verification signal is based on cross-seed consistency rather than the reconstruction quality of a prescribed image, and the generated batches retain consistent content under the same obfuscations.
This highlights the stability of the collapse phenomenon induced by our optimized embeddings, ensuring robust performance even under challenging conditions.

\subsection{Black-box API-only Verification}  \label{sec:black_exp}
We next consider the more restrictive API-only setting, where fingerprint conditions must be submitted as text prompts and verification relies solely on the returned images.

\input{figs/bb_unique}

\noindent
\textbf{Uniqueness Analysis.}  
We evaluate the distinctiveness of our method compared to TVN in Fig.~\ref{fig:bb_unique}. 
Pairwise cross-model verification should yield low scores only for matched fingerprint--model pairs along the main diagonal.
The results reveal that TVN suffers from a false positive (top-right cell), where SD 2.1 is misidentified as SD 1.4. 
In contrast, our method exhibits clearer diagonal separation, thereby ensuring reliable model identification.

\input{tables/bb_robust}

\input{tables/bb_robust_ft}

\noindent
\textbf{Robustness Analysis.}  
Before deployment, adversaries may apply obfuscation techniques, such as pruning, quantization, or fine-tuning, to modify model parameters and evade fingerprint verification. In Table~\ref{tab:bb_robust} and Table~\ref{tab:bb_ft}, we present the verification results under these attacks. 
The results show that TVN exhibits poor robustness, frequently misidentifying SD 2.1 as SD 1.4 across all attack types. 
Moreover, pruning SD 1.4 leads to the same false positive, which only ceases at $30\%$ pruning. This is likely because the model's ability to comprehend prompts and generate corresponding images is severely degraded at higher pruning levels. 
In contrast, our method remains effective across all attack types and intensities. 
This demonstrates that the proposed collapsed prompts induce a highly stable and consistent generation, where minor parameter modifications are insufficient to disrupt the established generation behavior. 
These findings validate the robustness of our approach under realistic adversarial conditions.

\input{tables/bb_adaptive_attack}

\noindent
\textbf{Adaptive Obfuscations.}  
We further consider scenarios where an adversary is aware of our fingerprinting method and attempts to bypass verification using adaptive attacks. 
Assuming that the verification query is detected (which is non-trivial as discussed in Sec.~\ref{sec:fp_verify}), the adversary may try to disrupt generation consistency. 
To evaluate this, we test four collapse-mitigation strategies: GPT rewriting, Random Token Addition (RTA)~\cite{somepalli2023understanding}, Embedding Optimization (EO)~\cite{wen2024detecting} and Sharpness-Aware Initialization for Latent Diffusion (SAIL)~\cite{jeon2025understanding} on SD 1.4. 
RTA inserts $R$ random tokens at arbitrary positions within the prompt, while EO and SAIL reduce the classifier-free guidance magnitude at the initial denoising step through prompt embedding or latent optimization. 
As shown in Table~\ref{tab:bb_aa}, our fingerprints remain effective against all obfuscation strategies, with $p_{\text{val}}$ consistently staying well below the threshold. 
Given the difficulty of detecting verification queries, adversaries are forced to apply these mitigations routinely. 
However, some interventions reduce semantic quality, while all introduce additional processing latency.
This demonstrates that our method has a stable collapse effect, ensuring robustness even against adaptive obfuscations.

\noindent
\textbf{Stealthiness Analysis.}  
To evade verification, an adversary may attempt to detect fingerprint prompts. 
A common detection approach involves analyzing prompt perplexity, as unnatural or garbled text often indicates abnormal queries. 
We evaluate stealthiness by measuring the average perplexity of the fingerprint prompts. Results in Table~\ref{tab:bb_stealthy} show that TVN's prompts exhibit extremely high perplexity (all exceeding $200$) due to their suffix-based design. 
As an adversary, one can further locate and remove the suspicious suffixes to evade verification. 
In contrast, our fingerprints retain the form of naturally occurring, human-readable prompts, resulting in substantially lower perplexity.

\input{tables/bb_stealthy}

\subsection{Efficiency}

To answer RQ4, we examine the computational cost of fingerprint construction.
Table~\ref{tab:cost} reports the per-fingerprint runtime of optimization-based construction procedures under their respective access settings.
Under pipeline access, our truncated embedding synthesis requires $25.03$ seconds, comparable to the $24.91$ seconds required by FingerInv.
For reference, TVN requires $369.57$ seconds to construct one API-compatible fingerprint through iterative prompt-suffix optimization.
These results show that our active fingerprint synthesis incurs practical construction overhead comparable to the pipeline-access baseline.

\input{tables/cost}

%% file: figs/cifar.tex
\begin{figure}[t] 
    \centering

    \newlength{\subfigheight}
    \setlength{\subfigheight}{0.12\textheight}
    
    \begin{subfigure}[b]{0.38\linewidth} 
        \centering
        \begin{minipage}[c][\subfigheight][c]{\linewidth}
            \centering
            \includegraphics[width=\linewidth,height=\subfigheight,keepaspectratio]{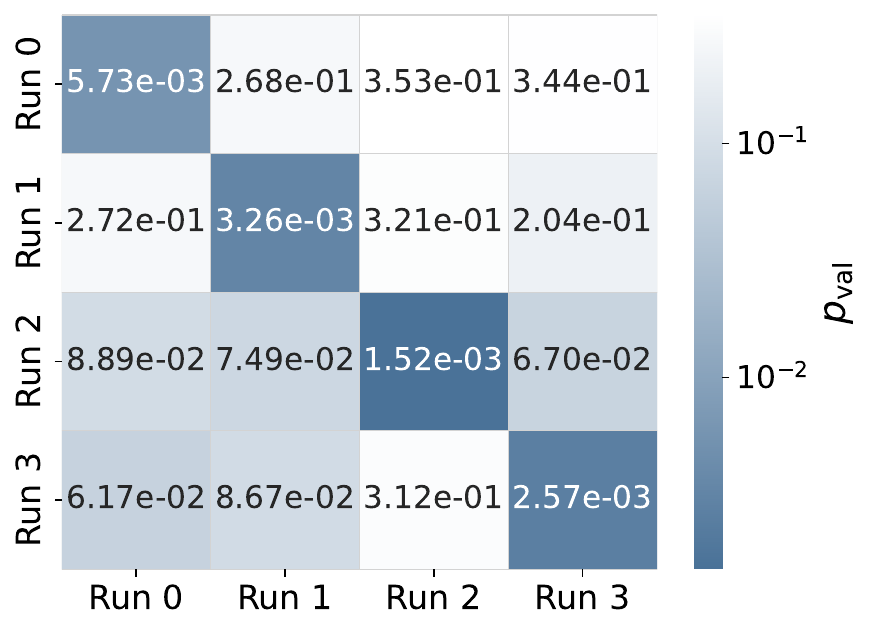}
        \end{minipage}
        \caption{Confusion matrix}
    \end{subfigure}
    \hfill 
    \begin{subfigure}[b]{0.58\linewidth}
        \centering
        \begin{minipage}[c][\subfigheight][c]{\linewidth}
            \centering
            \includegraphics[width=\linewidth,height=\subfigheight,keepaspectratio]{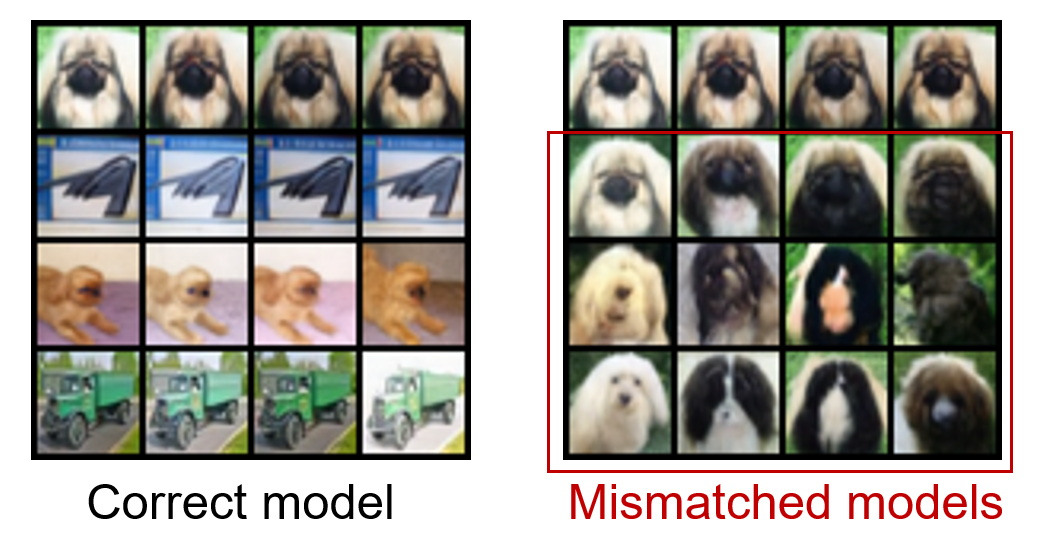}
        \end{minipage}
        \caption{Representative generations}
    \end{subfigure}
    
    \caption{Controlled validation of fingerprint uniqueness across four independently trained conditional DDPMs. (a) Cross-model verification matrix. Rows represent the source models of the fingerprint sets, columns represent the queried models, and lower $p_{\mathrm{val}}$ indicate stronger matches. (b) Representative cross-seed generations. The left panel shows one fingerprint condition from each model evaluated on its corresponding source model. The right panel applies the first fingerprint condition from the left panel to all four models, producing consistent outputs on the matched source model and diverse outputs on the three mismatched models.}
    \label{fig:cifar}
\end{figure}

%% file: figs/wb_unique.tex
\begin{figure}[t] 
    \centering
    
    \begin{subfigure}[b]{0.48\linewidth} 
        \centering
        \includegraphics[width=\linewidth]{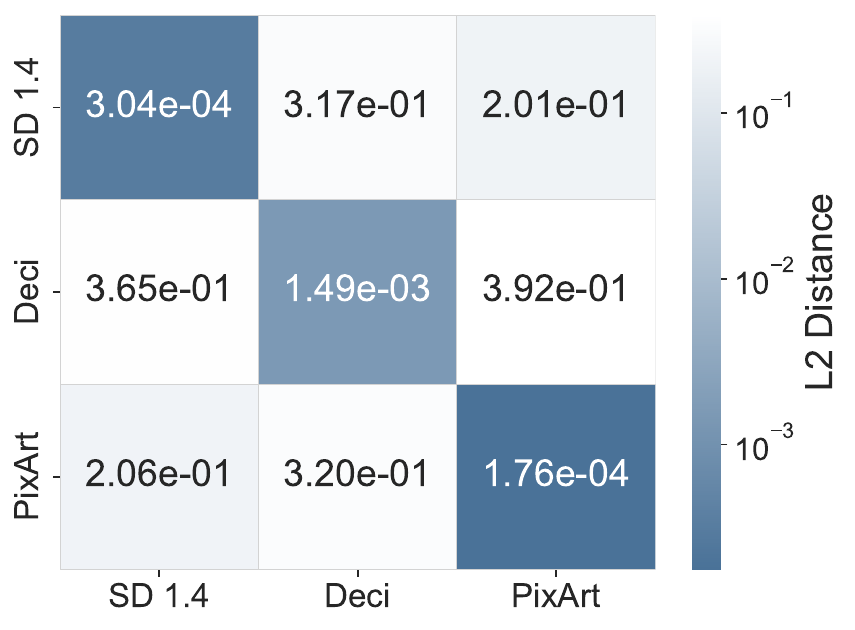}
        \caption{FingerInv~\cite{teng2025fingerprinting}} 
    \end{subfigure}
    \hfill 
    \begin{subfigure}[b]{0.48\linewidth}
        \centering
        \includegraphics[width=\linewidth]{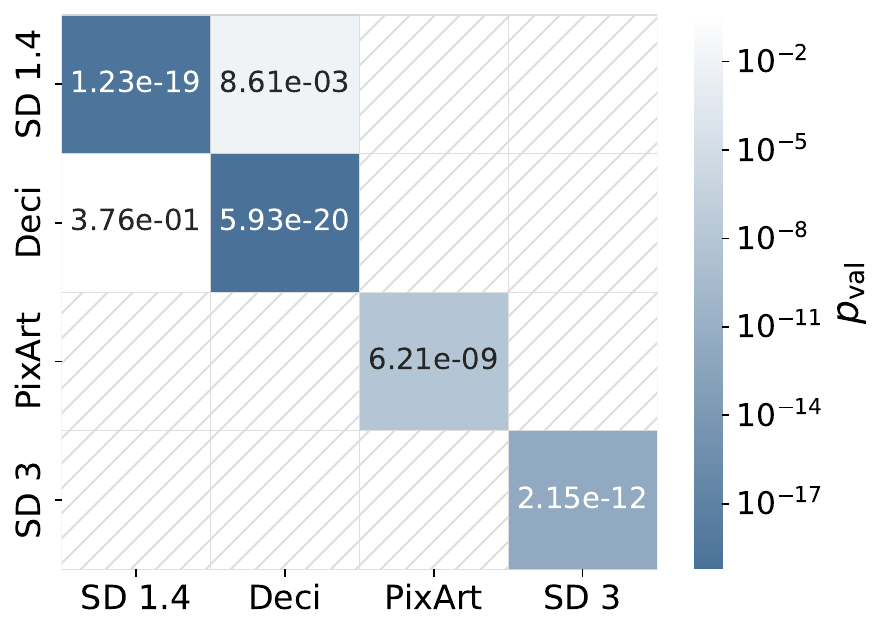}
        \caption{Ours} 
    \end{subfigure}

    \caption{Confusion matrices (white-box). Blue indicates higher likelihood that the suspect model is identified as the corresponding source model. Due to architectural differences, our fingerprints of PixArt and SD 3 cannot be used with other models, resulting in empty cells.}
    \label{fig:wb_unique}
\end{figure}

%% file: figs/wb_robust_cases.tex
\begin{figure}[t]
    \centering
    \includegraphics[width=\linewidth]{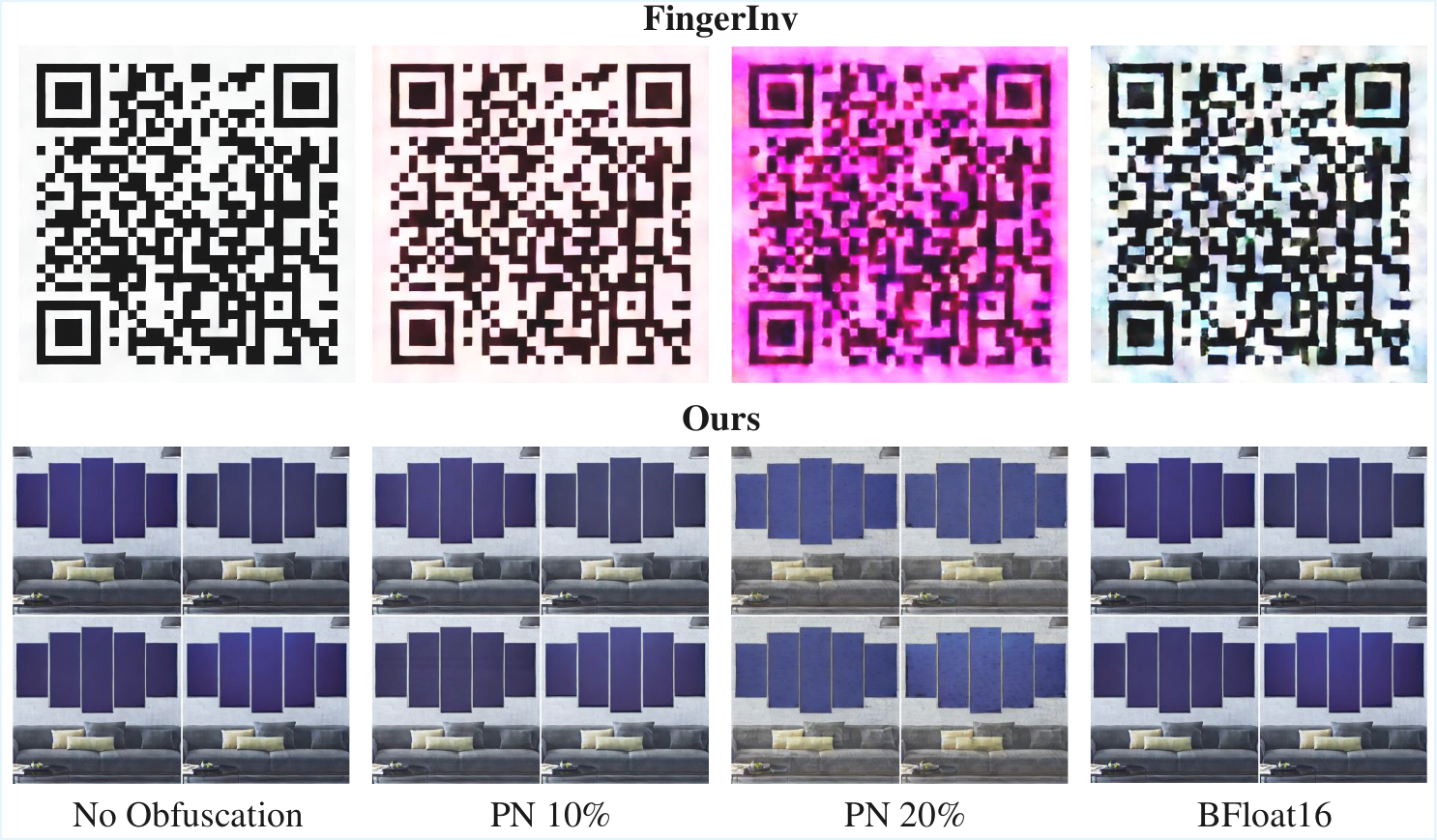}
    \caption{Case study for robustness analysis (white-box). FingerInv demonstrates degradation in the quality of generated QR codes under strong obfuscations, which may hinder verification. In contrast, our fingerprint relies on the consistency within the generated batch, maintaining strong robustness.}
    \label{fig:wb_robust_cases}
\end{figure}

%% file: figs/bb_unique.tex
\begin{figure}[t] 
    \centering
    
    \begin{subfigure}[b]{0.45\linewidth} 
        \centering
        \includegraphics[width=\linewidth]{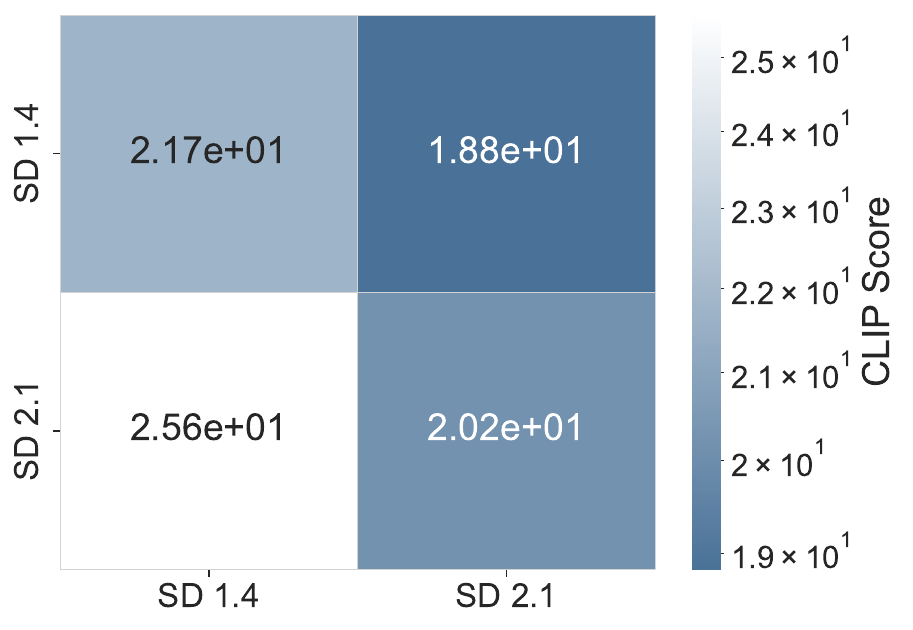} 
        \caption{TVN~\cite{guo2025one}} 
    \end{subfigure}
    \hfill 
    \begin{subfigure}[b]{0.45\linewidth}
        \centering
        \includegraphics[width=0.96\linewidth]{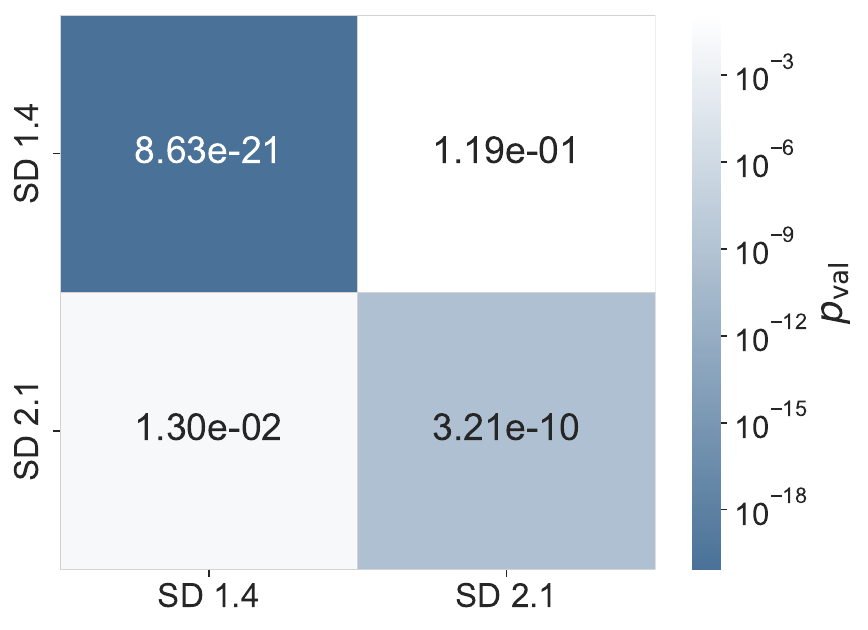}
        \caption{Ours} 
    \end{subfigure}
    
    \caption{Confusion matrices (black-box). Scores are displayed on a logarithmic scale, with lower values indicating stronger matches. Our method (b) shows clear unique identification for each suspect model, while TVN (a) exhibits significant confusion among different models.}
    \label{fig:bb_unique}
\end{figure}

%% file: tables/bb_robust.tex
\begin{table}[t]
\centering
\caption{Robustness (black-box, quantization and pruning). Green and red cells indicate successful and failed verification, respectively. $\dagger$ indicates false alarm.}
\label{tab:bb_robust}

\newcommand{\goodcell}[1]{\cellcolor[HTML]{DFF5DF}#1}
\newcommand{\badcell}[1]{\cellcolor[HTML]{F8D7DA}#1}

\resizebox{\linewidth}{!}{
\begin{tabular}{lcccc}
\toprule
\multirow{2}{*}{\textbf{Obfuscation}}
& \multicolumn{2}{c}{\textbf{TVN~\cite{guo2025one}} {\scriptsize (CLIP score)}}
& \multicolumn{2}{c}{\textbf{Ours} {\scriptsize ($p_{\mathrm{val}}$)}} \\
\cmidrule(lr){2-3} \cmidrule(lr){4-5}
& \textbf{SD 1.4}
& \textbf{SD 2.1}
& \textbf{SD 1.4}
& \textbf{SD 2.1} \\
\midrule

Quant. BF16
& \goodcell{21.8}
& \badcell{20.0$^\dagger$}
& \goodcell{\num{9.77e-21}}
& \goodcell{\num{3.44e-10}} \\

Quant. INT8
& \goodcell{21.7}
& \badcell{21.4$^\dagger$}
& \goodcell{\num{1.05e-20}}
& \goodcell{\num{2.74e-10}} \\

Quant. FP4
& \goodcell{20.5}
& \badcell{19.8$^\dagger$}
& \goodcell{\num{2.05e-20}}
& \goodcell{\num{1.49e-7}} \\

Prune 10\%
& \badcell{22.0$^\dagger$}
& \badcell{20.6$^\dagger$}
& \goodcell{\num{1.19e-20}}
& \goodcell{\num{5.00e-10}} \\

Prune 20\%
& \badcell{20.5$^\dagger$}
& \badcell{19.8$^\dagger$}
& \goodcell{\num{8.08e-20}}
& \goodcell{\num{6.35e-10}} \\

Prune 30\%
& \goodcell{24.3}
& \badcell{18.9$^\dagger$}
& \goodcell{\num{2.98e-16}}
& \goodcell{\num{6.26e-10}} \\

\bottomrule
\end{tabular}
}
\end{table}

%% file: tables/bb_robust_ft.tex
\begin{table}[t]
\centering
\caption{Robustness (black-box, fine-tuning). Green and red cells indicate successful and failed verification, respectively. $\dagger$ indicates false alarm.}
\label{tab:bb_ft}

\newcommand{\goodcell}[1]{\cellcolor[HTML]{DFF5DF}#1}
\newcommand{\badcell}[1]{\cellcolor[HTML]{F8D7DA}#1}

\resizebox{0.7\linewidth}{!}{
\begin{tabular}{lcc}
\toprule
\textbf{Model}
& \textbf{TVN~\cite{guo2025one}} {\scriptsize (CLIP score)}
& \textbf{Ours} {\scriptsize ($p_{\mathrm{val}}$)} \\
\midrule

SD 1.5
& \goodcell{\num{21.4}}
& \goodcell{\num{1.55e-20}} \\

Deliberate
& \badcell{\num{17.8}$^\dagger$}
& \goodcell{\num{6.50e-8}} \\

Realistic
& \goodcell{\num{21.9}}
& \goodcell{\num{5.00e-21}} \\

\bottomrule
\end{tabular}
}
\end{table}

%% file: tables/bb_adaptive_attack.tex
\begin{table}[t]
\centering
\caption{Robustness (black-box, adaptive obfuscations).}
\resizebox{0.9\linewidth}{!}{
\begin{tabular}{lcccc}
\toprule
                                      & {$p_{\text{val}}$ ($\downarrow$)} & {CLIP Score ($\uparrow$)} & Time (s) ($\downarrow$)                    & {Success?} \\ \midrule
{No Obfuscation} & $8.63 \times 10^{-21}$                  & {31.41}      & {1.744} & {\color[HTML]{77DD77} \CheckmarkBold}                       \\
{GPT Rewriting}  & $1.34 \times 10^{-15}$                  & 31.75                             & {2.590} & {\color[HTML]{77DD77} \CheckmarkBold}                       \\
{RTA~\cite{somepalli2023understanding}}            & $6.93 \times 10^{-16}$                  & {29.00}      & {1.753} & {\color[HTML]{77DD77} \CheckmarkBold}                       \\
{EO~\cite{wen2024detecting}}             & $2.33 \times 10^{-6}$                   & {28.21}      & {2.466} & {\color[HTML]{77DD77} \CheckmarkBold}                       \\
{SAIL~\cite{jeon2025understanding}}           & $2.24 \times 10^{-6}$                   & {30.25}      & {3.426} & {\color[HTML]{77DD77} \CheckmarkBold}                       \\ \bottomrule
\end{tabular}
}
\label{tab:bb_aa}
\end{table}

%% file: tables/bb_stealthy.tex
\begin{table}[t]
\centering
\caption{Perplexity ($\downarrow$) of fingerprint prompts (black-box).}
\resizebox{0.4\linewidth}{!}{
\begin{tabular}{lcc}
\toprule
Method & SD 1.4 & SD 2.1 \\ \midrule
TVN~\cite{guo2025one}                         & 294.96 & 243.83 \\
Ours                        & 41.85  & 51.32  \\ \bottomrule
\end{tabular}
}
\label{tab:bb_stealthy}
\end{table}

%% file: tables/cost.tex
\begin{table}[t]
\centering
\caption{Optimization time (s/prompt).}
\resizebox{0.6\linewidth}{!}{
\begin{tabular}{lccc}
\toprule
Method   & TVN~\cite{guo2025one}                  & FingerInv~\cite{teng2025fingerprinting}            & Ours                 \\  \midrule
Time ($\downarrow$)  & 369.57               & 24.91                & 25.03          
         \\ \bottomrule
\end{tabular}
}
\label{tab:cost}
\end{table}

%% file: tex/conclusion.tex
\section{Conclusion}

In this work, we formulate \emph{collapsed generation}, characterized by unusually high cross-seed consistency under particular input conditions, as a model-intrinsic behavioral signature for non-invasive ownership verification of T2I diffusion models.
Under pipeline access, we synthesize model-specific continuous embedding fingerprints through truncated optimization.
Under API-only access, we construct natural-prompt fingerprints by screening collapse-prone candidates.
Both fingerprint representations are evaluated through a unified source-referenced statistical test that determines whether a suspect model reproduces the characteristic collapse response of the source model.

A controlled study of independently trained conditional DDPMs and evaluations across representative U-Net-, DiT-, and flow-matching-based T2I models show clear separation between matched and mismatched fingerprint--model pairs, supporting fingerprint \emph{uniqueness}.
The fingerprints remain verifiable under most evaluated fine-tuned derivatives, model modifications, and adaptive query-time interventions, demonstrating \emph{robustness} across the considered conditions.
In the API-only setting, the natural fingerprint prompts remain human-readable and exhibit substantially lower perplexity than adversarially optimized prompt suffixes, supporting \emph{stealthier} service-level verification.
Meanwhile, truncated embedding synthesis maintains construction cost comparable to the pipeline-access baseline.
Together, these results establish collapsed generation as a practical source of intrinsic behavioral evidence for diffusion model ownership verification across checkpoint- and service-level disputes.

%% file: tex/supplements.tex
\input{tex/algorithm}

\section{Experiment Settings}  \label{sec:apdx_exp_setting}
\subsection{Datasets}
\begin{itemize}
    \item \textbf{Black-box prompts:} For SD 1.4 and SD 2.1, we use $500$ prompts each from~\cite{webster2023reproducible}. From these collections we pick prompts that produce highly collapsed generations to form the fingerprint set $\mathcal{P}_f$, following the filtering procedure described in Sec.~\ref{sec:fp_construct}.
    \item \textbf{White-box prompts:} For SD 1.4, we initialize with the template verbatim prompts from~\cite{webster2023reproducible}. For Deci and PixArt, we first apply the black-box prompt extraction procedure to the $1,000$ prompts to identify candidates exhibiting collapsed generations, then sample several of these candidates as initializations. We also evaluate that using low-diversity prompts and random initialization can also converge to effective fingerprints, as shown in Appendix~\ref{sec:further_analysis}.
    \item \textbf{GPT prompts:} For Fig.~\ref{fig:density} (a) and normal-generation modeling in the main experiments, we use GPT-4 to generate $50$ prompts in plain text. 
\end{itemize}

\subsection{Models}
All pretrained diffusion model checkpoints are obtained from HuggingFace~\cite{hugging}, including: 
\begin{itemize}
    \item \textbf{Target models:} 
    SD 1.4 (\path{CompVis/stable-diffusion-v1-4}),
    SD 2.1 (\path{stabilityai/stable-diffusion-2-1}),
    Deci (\path{Deci/DeciDiffusion-v1-0}),
    SD 3 (\path{stabilityai/stable-diffusion-3-medium-diffusers}),
    and PixArt (\path{PixArt-alpha/PixArt-XL-2-512x512}).
    \item \textbf{Fine-tune models:}
    SD 1.5 (\path{runwayml/stable-diffusion-v1-5}),
    Deliberate (\path{XpucT/Deliberate}),
    and Realistic (\path{SG161222/Realist_Vision_V2.0}).
\end{itemize}

\subsection{Implementation Details}
\noindent
\textbf{Baselines.}
\begin{itemize}
    \item \textbf{FingerInv~\cite{teng2025fingerprinting}:}
    We adopt the pipeline from the official open-source implementation provided by~\cite{teng2025fingerprinting}. Specifically, a QR code image of size $32\times32$ is used to obtain a model-specific DDPM latent code via FingerInv. This latent code is then applied to a suspect model to perform denoising. We subsequently examine whether the final generated image can still be successfully scanned. The quantitative measure is the L2 distance between the original and recovered QR codes. Note that this method is not applicable to flow-based models such as SD 3, which do not utilize DDPM-based sampling.
    \item \textbf{TVN~\cite{guo2025one}:} 
    This method identifies non-transferable adversarial prompts as fingerprints.
    Following the experimental setup in~\cite{guo2025one}, we start with $10$ initial prompts provided in the original work. For each tested model, we optimize a $5$-token prompt suffix using the NSGA-II black-box optimization algorithm. During optimization, the target model is treated as the positive example, while substitute models and a pretrained CLIP model serve as negative examples. The best-performing optimized prompt is selected as the fingerprint prompt for the target model.
    Subsequently, we run the generation with $10$ different random seeds and compute $\mu+3\sigma$ of the CLIP scores as the threshold.
\end{itemize}

\noindent
\textbf{Adaptive obfuscations.}
\begin{itemize}
    \item \textbf{GPT rewriting:}
    We employ GPT-4o to rewrite the original prompts using the following instruction:  \textit{"You are a prompt enhancer. For the prompt that will be input into the text-to-image model below, you can make some rewrites and optimizations to improve the image quality, but you must ensure that the semantics remain unchanged. Prompt to be rewritten: \{prompt\}}.
    \item \textbf{RTA~\cite{somepalli2023understanding}:} 
     Following the implementation in~\cite{wen2024detecting}, we randomly add four tokens for each prompt using the pipeline’s tokenizer.
    \item \textbf{EO~\cite{wen2024detecting}:}
    Following the official source code’s approach, we optimize the text embeddings for $10$ iterations at the first denoising step to reduce the classifier-free guidance magnitude, setting $l_{target}=1$.
    \item \textbf{SAIL~\cite{jeon2025understanding}:} 
    Following the original settings, we optimize the initial noise using an optimization threshold of $8.2$ and a maximum of $10$ iterations to steer the generation starting point away from the sharp regions described in the paper.
\end{itemize}

\noindent
\textbf{Our method.}
Our method consists of two verification settings: white-box and black-box. 
In the white-box setting, we apply the original text encoder of each diffusion model as the text encoder for prompt embedding extraction. 
In the black-box setting, we directly use the pipeline of each model for generation. 
We use different hyperparameters for different models due to their varying architectures and parameter scales.
We detail the hyperparameters in Table~\ref{tab:wb_hyperparams}.

\begin{table}[t]
\centering
\caption{Hyperparameters for white-box fingerprint optimization. }
\begin{tabular}{lcccc}
\toprule
 & SD 1.4 & Deci & PixArt & SD 3\\
\midrule
Truncated step $t_\text{trunc}$ & 5 & 5 & 10 &10\\
Learning rate $\eta$ & 0.01 & 0.01 & 0.002 & 0.01\\
Optimization iters $I$ & 15 & 30 & 8 & 60 \\
\bottomrule
\end{tabular}
\label{tab:wb_hyperparams}
\end{table}

\noindent
\textbf{Metrics.}
\begin{itemize}
    \item \textbf{SSCD:} We use the SSCD~\cite{pizzi2022self} model (\path{sscd_disc_large}) provided in the GitHub repository of the original paper for collapsed generation verification.
    \item \textbf{CLIP Score:} We use the \path{openai/clip-vit-base-patch32} model from the \path{transformers} library to measure the semantic similarity between generated images and text prompts on embeddings. 
\end{itemize}

\noindent
\textbf{Computational resources.}
All the experiments are conducted on a machine equipped with $8$ NVIDIA RTX 4090 GPUs, each with $24$ GB of VRAM, except for white-box fingerprint optimization on PixArt, which is performed on a single NVIDIA A100 GPU with $80$ GB of VRAM.

\section{Additional Experiment Results}   \label{sec:apdx_additional_results}

\subsection{Robust Analysis in the White-box Setting}

\input{tables/wb_robust}

\input{tables/wb_robust_ft}
Table \ref{tab:wb_robust} and \ref{tab:wb_robust_ft} present a comprehensive robustness analysis under the white-box setting. The results demonstrate that both our method and FingerInv remain robust against the majority of obfuscation techniques, including quantization, pruning, and fine-tuning.

However, both fingerprinting methods fail on lightweight architectures such as Deci. This failure is attributed to the severe performance degradation caused by $30\%$ pruning, which renders the model incapable of generating semantically meaningful images. Notably, such drastic obfuscation is unlikely to be applied by attackers in realistic threat scenarios.

\subsection{Ablation Study}

\input{figs/bb_ablation}

\input{figs/wb_ablation}

We investigate the impact of varying the number of fingerprint prompts (embeddings) $M$ and number of generations $K$ on ownership verification. 
We evaluate the $p_{\text{val}}$ derived from average pairwise similarity scores using $M=\{2,4,8\}$ and $K=\{2,4,8\}$. 
The ablation results for black-box and white-box settings are presented in Fig.~\ref{fig:bb_ablation} and Fig.~\ref{fig:wb_ablation}, respectively. 
While increasing the number of prompts/embeddings to $8$ yields the optimal $p_{\text{val}}$, the performance gain is marginal. 
Conversely, using $4$ seeds is sufficient to maintain a significantly low $p_{\text{val}}$, since inducing collapsed generations over $4$ random initializations is already distinctive enough for ownership verification.
To balance performance with the query budget, we select $M=4$ prompts/embeddings and $K=4$ seeds as the default configuration.

\subsection{Case Study}

\input{figs/unique_cases}

\input{figs/bb_robust_cases_appendix}

\input{figs/wb_robust_cases_appendix}

Fig.~\ref{fig:unique_cases} provides qualitative examples for the black-box and white-box uniqueness analysis. Notably, only the image groups along the main diagonal maintain a high degree of visual consistency, while off-diagonal entries exhibit varying levels of significant diversity. This validates the strong discriminative ability of our proposed average pairwise similarity score.

Furthermore, Fig.~\ref{fig:bb_robust_cases_appendix} and \ref{fig:wb_robust_cases_appendix} demonstrate the robustness of our method in both settings. Taking SD 1.4 as the target model, the generated images retain high similarity scores even when subjected to standard obfuscations (quantization, pruning, fine-tuning) or adaptive attacks (e.g., RTA, EO). Moreover, such consistency is not affected even when the quality of the generated images is degraded. These results confirm that our method is highly robust and applicable to diverse real-world threat scenarios.

\section{Further Analysis} \label{sec:further_analysis}

\subsection{Scalability of Black-Box Fingerprints in Recent Models}
Following early diffusion models, recent models (e.g., SD 3~\cite{esser2024scaling}) have heavily employed rigorous data deduplication pipelines to explicitly suppress the phenomenon of collapsed generation. 
One might question the scalability of finding natural collapsed text prompts in such heavily deduplicated models. 
However, while deduplication effectively mitigates simple memorization caused by exact data duplication, recent studies~\cite{ross2025geometric,bonnaire2025why} reveal that collapsed generation is also driven by other factors, including intrinsic data characteristics (e.g., outlier samples) and training dynamics (e.g., overtraining).
This ensures that a sparse set of natural candidates inevitably persists even in highly curated datasets. 
Because these natural collapsed prompts carve deep, specific basins in the loss landscape during the extensive training phase, they inherently exist as discrete text tokens.
Importantly, our framework only requires a minimal set of fingerprints (e.g., $|\mathcal{P}_f| = 4$) for reliable verification, making this sparse availability more than sufficient. 
Furthermore, model owners can efficiently capture these rare collapsed prompts ``on the fly'' during training via loss monitoring~\cite{bonnaire2025why} as illustrated in Sec.~\ref{sec:fp_construct}, bypassing the need for expensive post-hoc scanning.

\subsection{Efficacy of White-Box Optimization on Recent Models}
While natural collapsed generation is reduced in recent models, our white-box optimization provides a proactive mechanism to deliberately induce such behaviors. 
Unlike black-box prompts that rely on naturally occurring collapsed generation stemming from training outliers and artifacts, white-box optimization actively forces the model's latent trajectory into a collapsed manifold via gradient descent in the continuous embedding space. 
As demonstrated in our experiments with PixArt and SD 3, we can still reliably optimize continuous embeddings that trigger collapsed generations in these recent models, ensuring the framework's long-term applicability.

\input{tables/dicrete_optim}

\subsection{Challenges in Discrete Fingerprint Optimization}
Given the success of continuous white-box optimization, an intuitive extension is to map these optimized embeddings back to discrete text tokens to enable purely black-box verification. 
However, we find such discretization highly non-trivial in practice. 

The fundamental reason is that collapsed modes reside in extremely narrow and precise regions within the latent space~\cite{ross2025geometric}. 
While natural black-box prompts can successfully trigger collapse because the model explicitly learned their discrete mappings during training, our white-box optimized embeddings locate artificial collapsed points. 
When applying post-hoc projection (mapping optimized embeddings to the nearest vocabulary tokens) or gradient-based discrete optimization (e.g., PEZ~\cite{wen2023hard} or GCG~\cite{zou2023universal}), the necessary discretization step severely disrupts the precise continuous coordinates required to stay within these narrow basins, resulting in a significant drop in SSCD scores (leading to $p_\text{val} > 10^{-4}$ in Table~\ref{tab:dicrete_optim}). 
Consequently, we maintain the white-box fingerprints in the continuous embedding space and leave the discretization of collapsed embeddings as a valuable direction for future work.

\input{tables/init}

\subsection{Different Initialization Strategies in White-Box Optimization}
We evaluate the impact of different initialization prompts on white-box optimization in Table~\ref{tab:init}. While warm starts based on collapsed prompts accelerate convergence, they are not strictly necessary. Low-diversity prompts generated by LLMs (e.g., pure textual) are also sufficient. Even with random initialization, the optimization successfully converges, although it requires more steps.

\section{Prompts Used in Visualization}

\subsection{Figure~\ref{fig:intro}}
\begin{itemize}
    \item \textit{The No Limits Business Woman Podcast}
\end{itemize}

\subsection{Figure~\ref{fig:density}}
\begin{itemize}
    \item \textbf{GPT Generated Prompt:} \textit{A glass teapot house in a foggy forest}
    \item \textbf{SD 1.4 Collapsed Prompt:} \textit{\textless i\textgreater The Long Dark\textless /i\textgreater  Gets First Trailer, Steam Early Access}
    \item \textbf{SD 2.1 Collapsed Prompt:} \textit{Pencil pleat curtain in collection Linen, fabric: 392-05}
\end{itemize}

\subsection{Figure~\ref{fig:overview}}
\begin{itemize}
    \item \textit{Lilah Teal Blue Area Rug by Andover Mills}
\end{itemize}

\subsection{Figure~\ref{fig:mem_vs_nonmem_steps}}
\begin{itemize}
    \item \textbf{Collapsed:} \textit{Thumbprintz Splatter No I Red Floor Pillow} 
    \item \textbf{Normal:} \textit{Happy woman in hat with shopping bags showing credit card}
\end{itemize}

\subsection{Figure~\ref{fig:unique_cases}}
\begin{itemize}
    \item \textbf{SD 1.4 (black-box):} \textit{\textless em\textgreater Bloodborne\textless\/em\textgreater Video: Sony Explains the Game's Procedurally Generated Dungeons} 
    \item \textbf{SD 2.1:} \textit{Pencil pleat curtains in collection Jupiter, fabric: 127-00}
    \item \textbf{SD 1.4 (white-box), Deci:} \textit{N/A}\footnote{White-box fingerprints are optimized directly in the embedding space and therefore have no human-readable text prompt representation.}
\end{itemize}

\subsection{Figure~\ref{fig:bb_robust_cases_appendix}}
\begin{itemize}
     \item \textit{\textless em\textgreater Bloodborne\textless\/em\textgreater Video: Sony Explains the Game's Procedurally Generated Dungeons}
\end{itemize}

%% file: tex/algorithm.tex
\section{Algorithms}  \label{sec:apdx_algorithm}

\begin{algorithm}
\caption{Continuous-Embedding Fingerprint Synthesis}
\label{alg:whitebox_opt}
\begin{algorithmic}[1]
\Require Initial prompt set
$\mathcal{P}_{\mathrm{init}}=\{p_i\}_{i=1}^{M}$,
source model $G_\theta$,
text encoder $E_{\mathrm{text}}$,
number of noise samples $K$,
number of truncated denoising steps $t_{\mathrm{trunc}}$,
number of optimization iterations $I$,
learning rate $\eta$
\Ensure Optimized fingerprint embedding set
$\mathcal{C}_f=\{c_i^*\}_{i=1}^{M}$

\State $\mathcal{C}_f \gets \emptyset$
\For{$i=1$ to $M$}
    \State $c_i^{(0)} \gets E_{\mathrm{text}}(p_i)$
    \Comment{Initialize the $i$-th embedding}
    \State Sample independent noises
    $\{x_T^{(k)}\}_{k=1}^{K}$

    \For{$r=0$ to $I-1$}
        \State Run the first $t_{\mathrm{trunc}}$ denoising steps with
        $c_i^{(r)}$ to obtain
        $\{x_{T-t_{\mathrm{trunc}}}^{(k)}(c_i^{(r)})\}_{k=1}^{K}$

        \State
        $\bar{x}_{T-t_{\mathrm{trunc}}}(c_i^{(r)})
        \gets
        \frac{1}{K}
        \sum_{k=1}^{K}
        x_{T-t_{\mathrm{trunc}}}^{(k)}(c_i^{(r)})$

        \State
        $\mathcal{L}_{\mathrm{Collapse}}(c_i^{(r)})
        \gets
        \frac{1}{K}
        \sum_{k=1}^{K}
        \left\|
        x_{T-t_{\mathrm{trunc}}}^{(k)}(c_i^{(r)})
        -
        \bar{x}_{T-t_{\mathrm{trunc}}}(c_i^{(r)})
        \right\|_2^2$

        \State
        $c_i^{(r+1)}
        \gets
        c_i^{(r)}
        -
        \eta
        \nabla_{c_i^{(r)}}
        \mathcal{L}_{\mathrm{Collapse}}(c_i^{(r)})$
    \EndFor

    \State $c_i^* \gets c_i^{(I)}$
    \State $\mathcal{C}_f \gets \mathcal{C}_f \cup \{c_i^*\}$
\EndFor

\State \Return $\mathcal{C}_f$
\end{algorithmic}
\end{algorithm}

\begin{algorithm}
\caption{Fingerprint Verification}
\label{alg:fp_verification_multi}
\begin{algorithmic}[1]
\Require Fingerprint set $\mathcal{F}=\{f_m\}_{m=1}^{M}$, where $\mathcal{F}$ is either $\mathcal{C}_f$ or $\mathcal{P}_f$; suspect model $G_{\theta'}$; source-model reference scores $\mathcal{U}_0=\{u_n\}_{n=1}^{N}$; number of generations $K$; decision threshold $\tau_f$
\Ensure Verification decision $d$ and statistical evidence $p_{\mathrm{val}}$

\State $\mu_0 \gets \frac{1}{N}\sum_{n=1}^{N}u_n$
\State $\sigma_0 \gets \sqrt{\frac{1}{N-1}\sum_{n=1}^{N}(u_n-\mu_0)^2}$
\State $\mathcal{S}_f \gets \emptyset$

\For{$m=1$ to $M$}
    \State Obtain $K$ independent outputs
    $\{x_0^{(k)}(f_m)\}_{k=1}^{K}$
    from $G_{\theta'}$ conditioned on $f_m$
    \Comment{$f_m$ is supplied as an embedding or a text prompt}

    \State Compute the cross-seed consistency
    $\bar{s}_{\theta'}(f_m,K)$ using Eq.~\ref{eq:similarity}

    \State $\mathcal{S}_f
    \gets
    \mathcal{S}_f
    \cup
    \{\bar{s}_{\theta'}(f_m,K)\}$
\EndFor

\State $\bar{s}_f
\gets
\frac{1}{M}
\sum_{s\in\mathcal{S}_f}s$

\State $t_f
\gets
\frac{\bar{s}_f-\mu_0}
{\sigma_0\sqrt{1+1/N}}$

\State $p_{\mathrm{val}}
\gets
1-F_{t_{N-1}}(t_f)$
\Comment{$F_{t_{N-1}}$ is the Student's $t$ CDF}

\If{$p_{\mathrm{val}}<\tau_f$}
    \State $d\gets\texttt{True}$
    \Comment{The suspect model matches the source fingerprint}
\Else
    \State $d\gets\texttt{False}$
    \Comment{The fingerprint match is not established}
\EndIf

\State \Return $(d,p_{\mathrm{val}})$
\end{algorithmic}
\end{algorithm}

%% file: tables/wb_robust.tex
\begin{table}[t]
\centering
\caption{Robustness (white-box, quantization and pruning). Green and red cells indicate successful and failed verification, respectively. }
\label{tab:wb_robust}

\newcommand{\goodcell}[1]{\cellcolor[HTML]{DFF5DF}#1}
\newcommand{\badcell}[1]{\cellcolor[HTML]{F8D7DA}#1}

\resizebox{0.95\linewidth}{!}{
\begin{tabular}{lcccc}
\toprule
\textbf{Obfuscation}
& \textbf{SD 1.4}
& \textbf{Deci}
& \textbf{PixArt}
& \textbf{SD 3} \\
\midrule

\multicolumn{5}{l}{\textbf{FingerInv~\cite{teng2025fingerprinting}} {\scriptsize ($L_2$ distance)}} \\
Quant. BF16
& \goodcell{\num{3.76e-3}}
& \goodcell{\num{2.84e-2}}
& \goodcell{\num{9.25e-5}}
& -- \\
Quant. INT8
& \goodcell{\num{4.52e-4}}
& \goodcell{\num{1.96e-3}}
& \goodcell{\num{6.31e-4}}
& -- \\
Quant. FP4
& \goodcell{\num{2.10e-4}}
& \goodcell{\num{4.12e-3}}
& \goodcell{\num{1.38e-3}}
& -- \\
Prune 10\%
& \goodcell{\num{1.63e-4}}
& \goodcell{\num{1.45e-2}}
& \goodcell{\num{1.97e-4}}
& -- \\
Prune 20\%
& \goodcell{\num{2.43e-4}}
& \goodcell{\num{1.11e-1}}
& \goodcell{\num{5.09e-3}}
& -- \\
Prune 30\%
& \goodcell{\num{1.30e-2}}
& \badcell{\num{3.41e-1}}
& \goodcell{\num{2.16e-2}}
& -- \\

\midrule

\multicolumn{5}{l}{\textbf{Ours} {\scriptsize ($p_{\mathrm{val}}$)}} \\
Quant. BF16
& \goodcell{\num{1.23e-19}}
& \goodcell{\num{6.11e-20}}
& \goodcell{\num{8.46e-8}}
& \goodcell{\num{8.38e-11}} \\
Quant. INT8
& \goodcell{\num{3.92e-19}}
& \goodcell{\num{6.23e-20}}
& \goodcell{\num{3.16e-7}}
& \goodcell{\num{1.39e-11}} \\
Quant. FP4
& \goodcell{\num{4.69e-18}}
& \goodcell{\num{4.89e-19}}
& \goodcell{\num{5.72e-9}}
& \goodcell{\num{1.53e-9}} \\
Prune 10\%
& \goodcell{\num{3.11e-19}}
& \goodcell{\num{8.37e-20}}
& \goodcell{\num{3.56e-10}}
& \goodcell{\num{2.51e-12}} \\
Prune 20\%
& \goodcell{\num{2.00e-20}}
& \goodcell{\num{2.96e-18}}
& \goodcell{\num{6.41e-6}}
& \goodcell{\num{6.08e-11}} \\
Prune 30\%
& \goodcell{\num{7.18e-9}}
& \badcell{\num{4.57e-2}}
& \goodcell{\num{7.08e-11}}
& \goodcell{\num{1.90e-11}} \\

\bottomrule
\end{tabular}
}
\end{table}

%% file: tables/wb_robust_ft.tex
\begin{table}[t]
\centering
\caption{Robustness (white-box, fine-tuning). Green and red cells indicate successful and failed verification, respectively. }
\label{tab:wb_robust_ft}

\newcommand{\goodcell}[1]{\cellcolor[HTML]{DFF5DF}#1}
\newcommand{\badcell}[1]{\cellcolor[HTML]{F8D7DA}#1}

\resizebox{0.65\linewidth}{!}{
\begin{tabular}{lcc}
\toprule
\textbf{Model}
& \textbf{FingerInv~\cite{teng2025fingerprinting}} {\scriptsize ($L_2$ distance)}
& \textbf{Ours} {\scriptsize ($p_{\mathrm{val}}$)} \\
\midrule

SD 1.5
& \goodcell{\num{5.47e-4}}
& \goodcell{\num{1.95e-18}} \\

Deliberate
& \goodcell{\num{2.66e-4}}
& \goodcell{\num{6.09e-14}} \\

Realistic
& \goodcell{\num{6.98e-4}}
& \goodcell{\num{1.18e-18}} \\

\bottomrule
\end{tabular}
}
\end{table}

%% file: figs/bb_ablation.tex
\begin{figure}[t] 
    \centering
    
    \begin{subfigure}[b]{0.48\linewidth} 
        \centering
        \includegraphics[width=\linewidth]{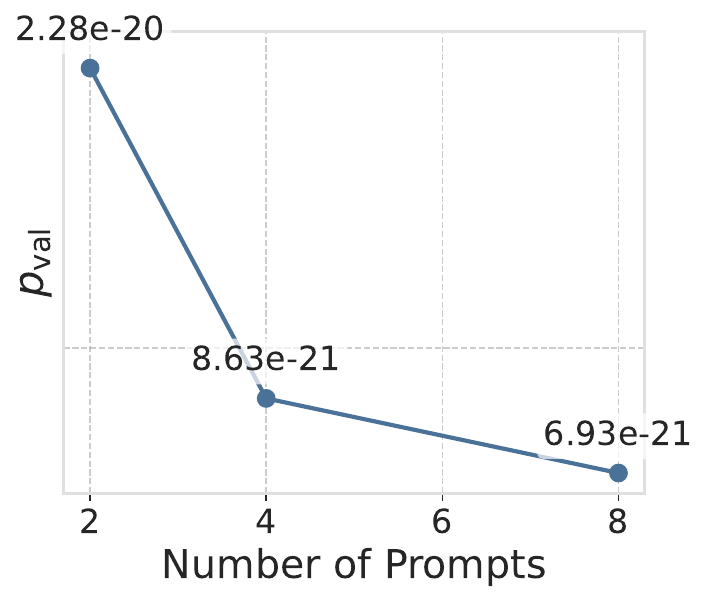} 
    \end{subfigure}
    \hfill 
    \begin{subfigure}[b]{0.48\linewidth}
        \centering
        \includegraphics[width=\linewidth]{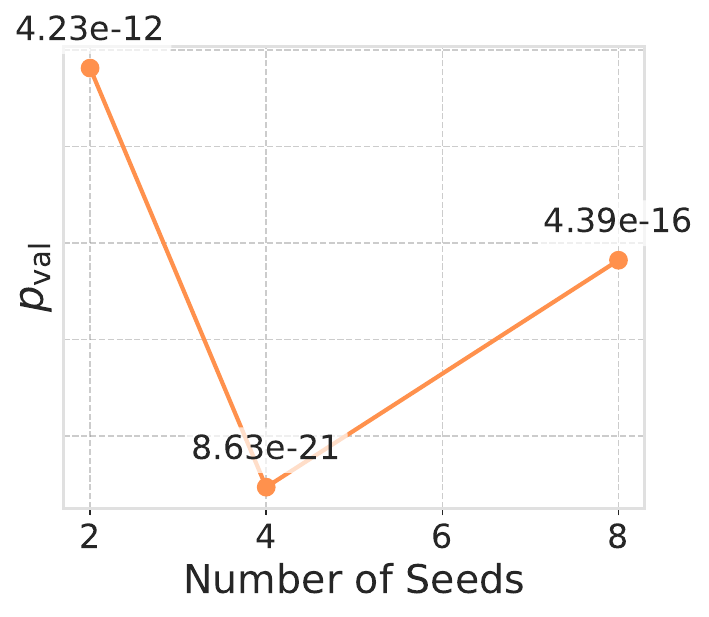}
    \end{subfigure}
    
    \caption{$p_{\text{val}}$ using different number of prompts/generations (black-box).}
    \label{fig:bb_ablation}
\end{figure}

%% file: figs/wb_ablation.tex
\begin{figure}[t] 
    \centering
    
    \begin{subfigure}[b]{0.48\linewidth} 
        \centering
        \includegraphics[width=\linewidth]{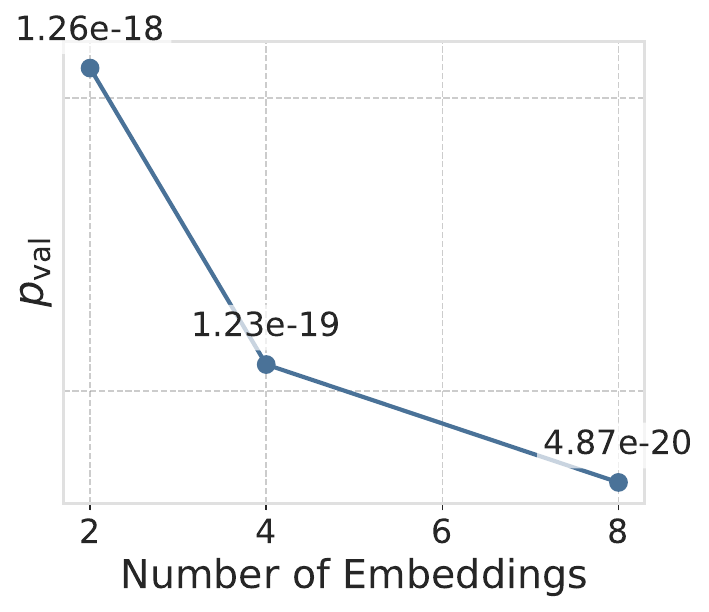} 
    \end{subfigure}
    \hfill 
    \begin{subfigure}[b]{0.48\linewidth}
        \centering
        \includegraphics[width=\linewidth]{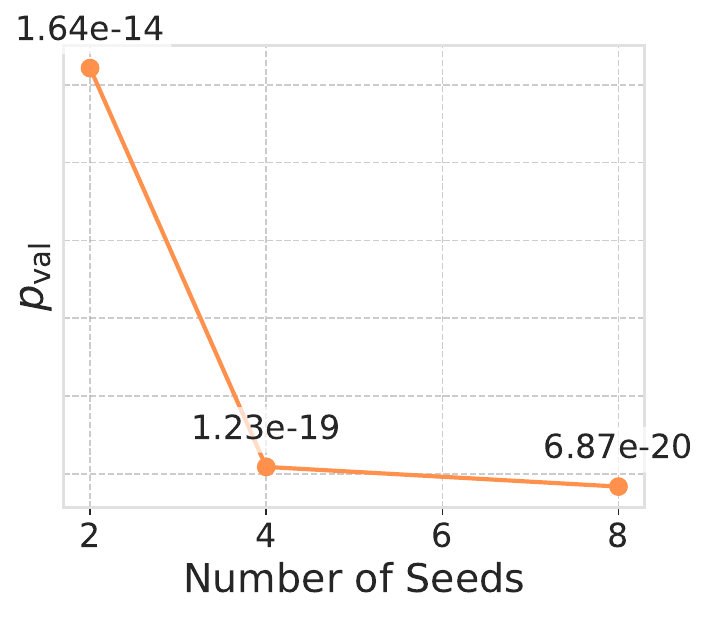}
    \end{subfigure}
    
    \caption{$p_{\text{val}}$ using different number of embeddings/generations (white-box).}
    \label{fig:wb_ablation}
\end{figure}

%% file: figs/unique_cases.tex
\begin{figure}[t]
    \centering
    \includegraphics[width=\linewidth]{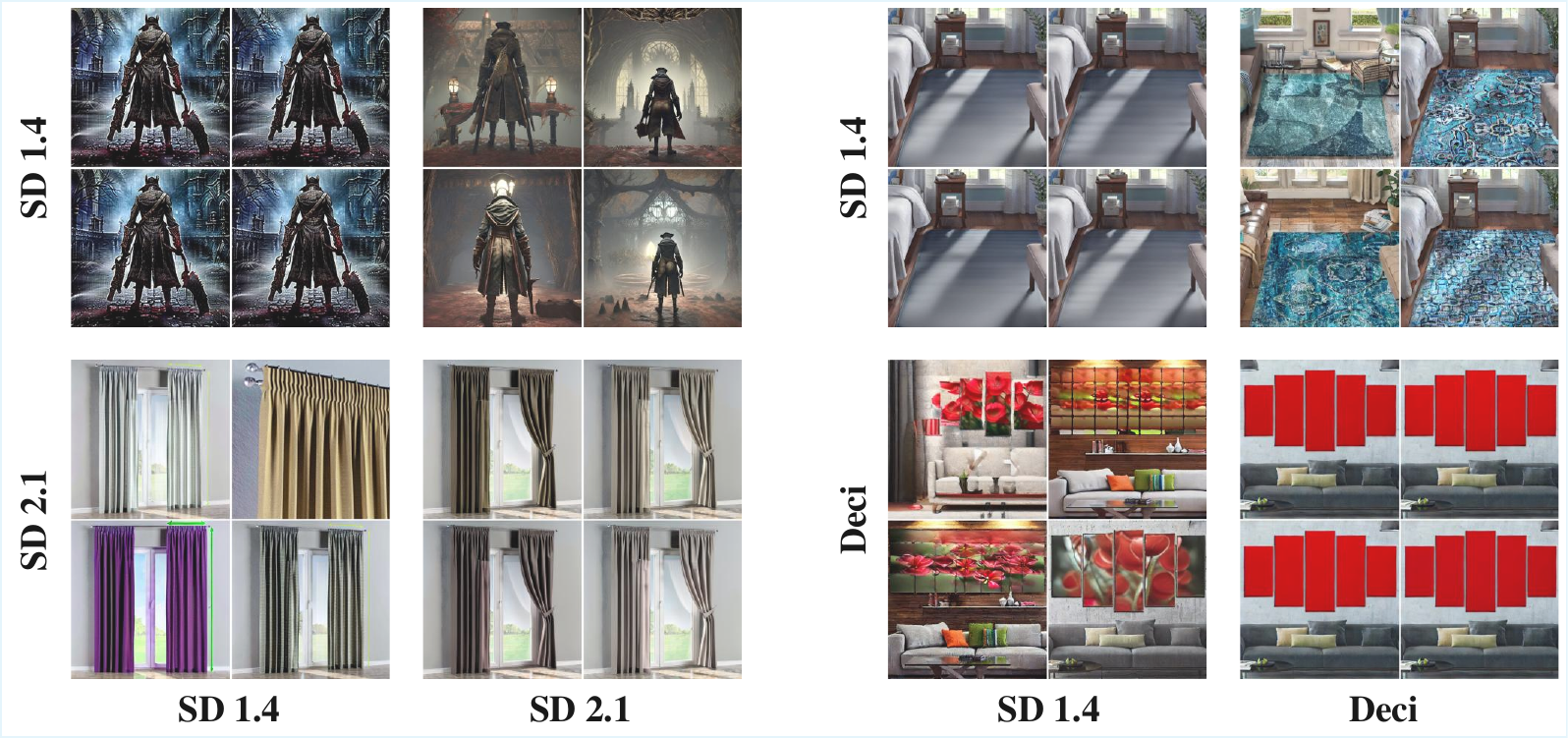}
    \caption{Case study for uniqueness analysis. Left: black-box, right: white-box.}
    \label{fig:unique_cases}
\end{figure}

%% file: figs/bb_robust_cases_appendix.tex
\begin{figure}[t]
    \centering
    \includegraphics[width=\linewidth]{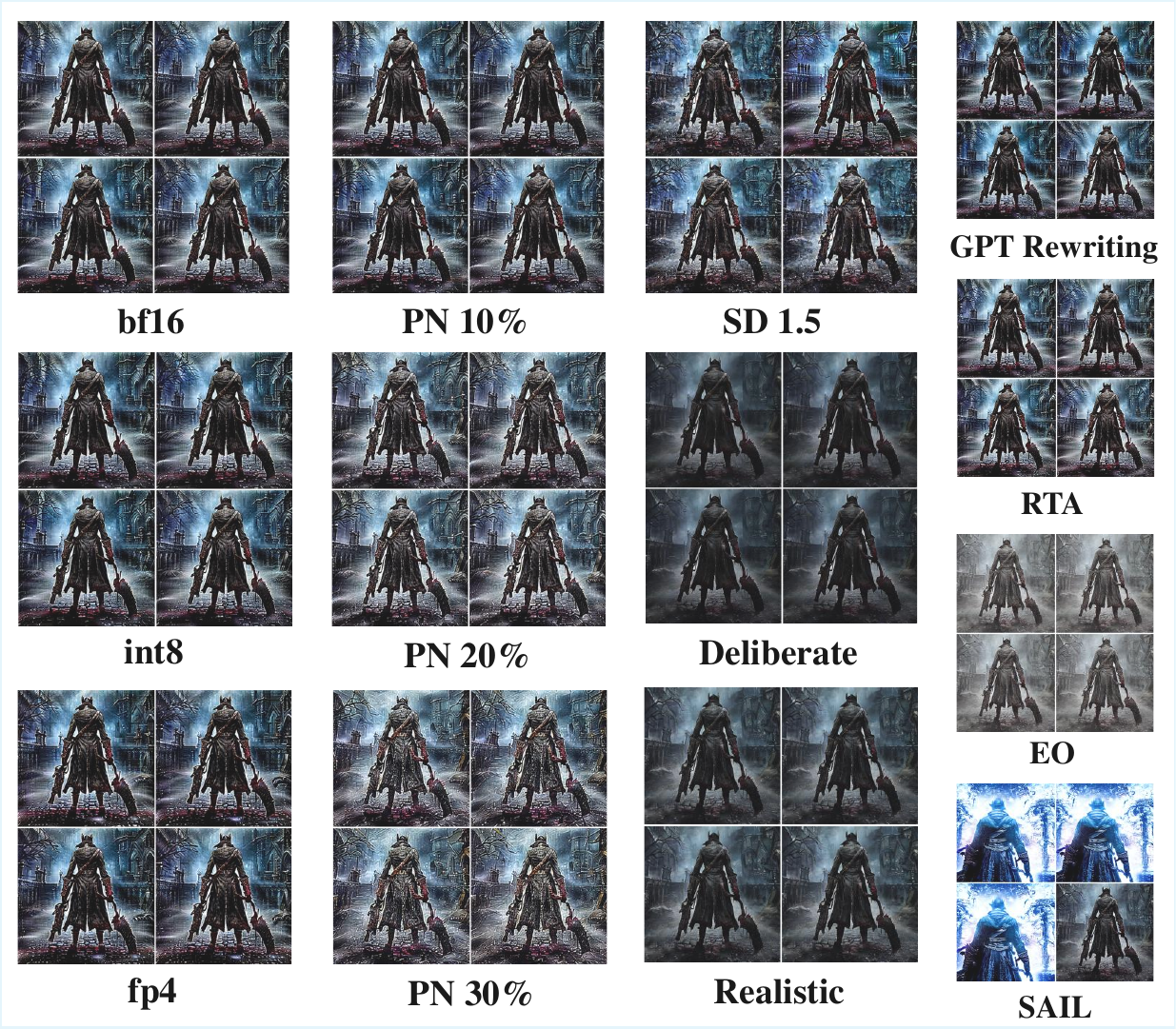}
    \caption{Case study for robustness analysis (SD 1.4, black-box).}
    \label{fig:bb_robust_cases_appendix}
\end{figure}

%% file: figs/wb_robust_cases_appendix.tex
\begin{figure}[t]
    \centering
    \includegraphics[width=\linewidth]{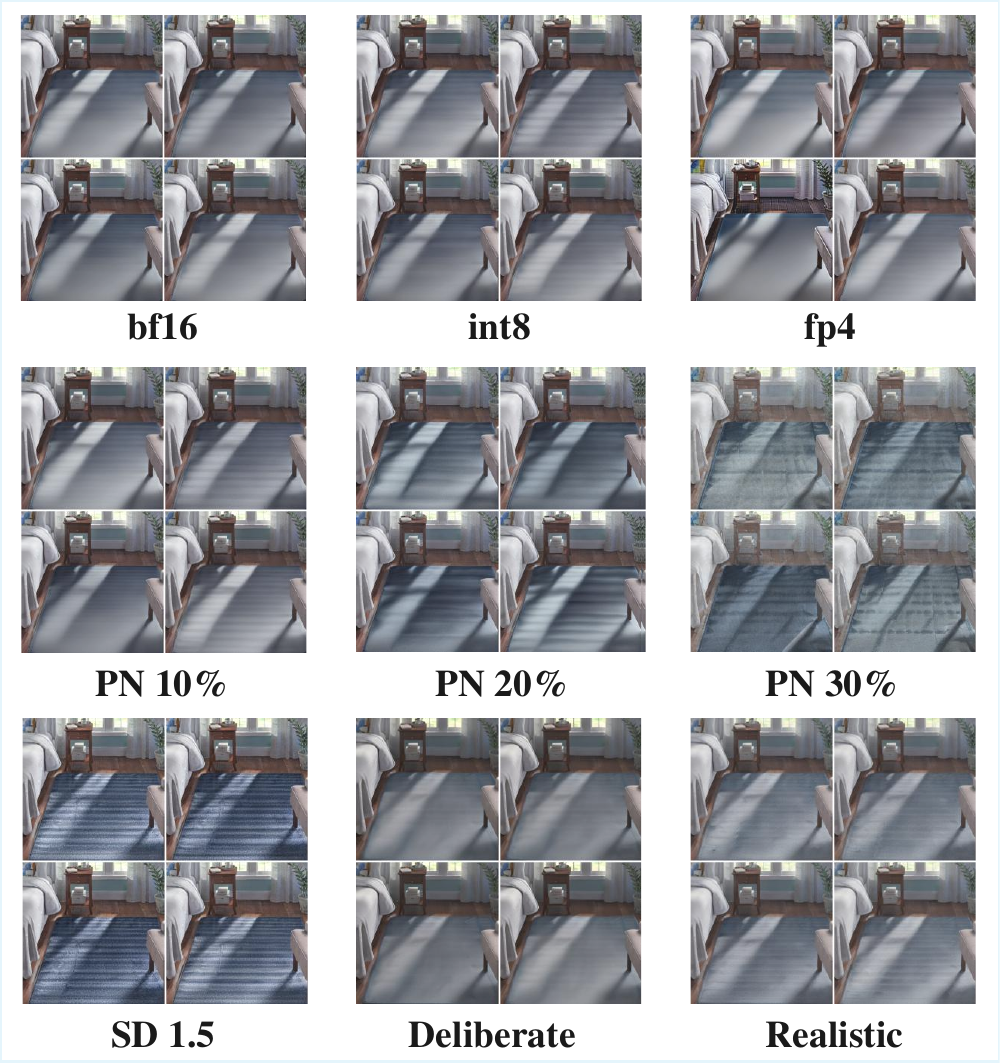}
    \caption{Case study for robustness analysis (SD 1.4, white-box).}
    \label{fig:wb_robust_cases_appendix}
\end{figure}

%% file: tables/dicrete_optim.tex
\begin{table}[t]
\centering
\caption{Embeddings obtained via discrete fingerprint optimization fail verification.}
\begin{tabular}{lccc}
\toprule
Strategy         & Post-hoc Proj.        & PEZ~\cite{wen2023hard}                   & GCG~\cite{zou2023universal}                   \\ \midrule
$p_{\text{val}}$ & $1.68 \times 10^{-4}$ & $4.23 \times 10^{-2}$ & $2.32 \times 10^{-4}$ \\
Success?         & {\color[HTML]{FF6961} \XSolidBrush}                     & {\color[HTML]{FF6961} \XSolidBrush}                     & {\color[HTML]{FF6961} \XSolidBrush}                     \\ \bottomrule
\end{tabular}
\label{tab:dicrete_optim}
\end{table}

%% file: tables/init.tex
\begin{table}[t]
\centering
\caption{Verification efficacy and time cost of different initialization strategies.}
\resizebox{0.8\linewidth}{!}{
\begin{tabular}{lccc}
\toprule
Init.                           & Random                   & Low Diversity            & Collapsed                \\ \midrule
$p_{\text{val}}$ ($\downarrow$) & $2.83 \times 10^{-6}$    & $4.26 \times 10^{-8}$    & $1.23 \times 10^{-19}$   \\
Time (s)                        & 47.77                   & 24.81                    & 25.03                    \\
{Success?} & {\color[HTML]{77DD77} \CheckmarkBold}  & {\color[HTML]{77DD77} \CheckmarkBold}  & {\color[HTML]{77DD77} \CheckmarkBold}  \\ \bottomrule
\end{tabular}
}
\label{tab:init}
\end{table}